\documentclass[structabstract]{aa}
\usepackage{graphicx}
\def\approxgt{\mathrel{\hbox{\rlap{\lower.55ex \hbox {$\sim$}}
        \kern-.3em \raise.4ex \hbox{$>$}}}}
\def\approxlt{\mathrel{\hbox{\rlap{\lower.55ex \hbox {$\sim$}}
        \kern-.3em \raise.4ex \hbox{$<$}}}}
\begin{document}
   \title{AGN/starburst connection in action: \\ the half million second
          RGS spectrum of NGC~1365}

   \author{M.Guainazzi
          \inst{1},
	  G.Risaliti,
	  \inst{2,3}
	  A.Nucita,
          \inst{1}
	  Junfeng~Wang,
	  \inst{2}
	  S.Bianchi,
	  \inst{4}
          R.Soria,
          \inst{5}
	  A.Zezas
          \inst{2,6}
          }

   \offprints{M.Guainazzi}

   \institute{$^1$European Space Astronomy Centre of ESA, P.O.Box 78,
              Villanueva de la Ca\~nada, E-28691 Madrid, Spain \\
	      \email{Matteo.Guainazzi@sciops.esa.int} \\
	      $^2$Harvard-Smithsonian Center for Astrophysics, Cambridge, MA
	      02318 \\
	      $^3$INAF-Osservatorio di Arcetri, Firenze, I-50125, Italy \\
	      $^4$Dipartimento di Fisica, Universit\`a degli Studi Roma Tre, via della Vasca Navale 84, I-00146 Roma, Italy \\
 	      $^5$Mullard Space Science Laboratory, University College London, Holmbury St Mary, Dorking, Surrey RH5 6NT, UK \\
	      $^6$Physics Department, University of Crete, P.O. Box 2208, GR-710 03, Heraklion, Crete, Greece
              }

   \date{Received ; accepted }

   \abstract{High-resolution X-ray observations in the imaging and spectral domain have recently opened a new window on active galactic nuclei (AGN) feedback onto the circumnuclear gas. Spectral diagnostics, as well as the remarkable morphological coincidence between [{\sc O iii}] and X-rays, point to AGN photoionisation as the dominant ionisation mechanism on scales as large as a few kpc.}
	    {In this paper we extend these studies to the nearby Seyfert~2 galaxy NGC~1365, known to host a circumnuclear ring of intense star formation at $\simeq$1.3~kpc from the nucleus. The main scope of this investigation is to study the connection between nuclear activity and star formation in nearby AGN.}
	    {We present a deep ($\simeq$5.8~days) 0.3--2~keV high-resolution spectrum of NGC~1365, collected with the reflection grating spectrometer (RGS) on board XMM-Newton.}
	    {The spectrum is dominated by strong recombination lines of He- and H-like transitions from carbon to silicon, as well as by L transitions from Fe{\sc xvii}. The continuum is strong, especially in the 10 to 20\AA\, range. Formal fits require two optically thin, collisionally ionised plasma components, with temperatures $\simeq$300 and $\simeq$640~eV. However, they leave the bulk of the forbidden components of the He-$\alpha$ O{\sc vii} and N{\sc vi} triplets unaccounted for. These features can be explained as being produced by photoionised gas. NGC~1365 is therefore the first obscured AGN, whose high-resolution X-ray spectrum requires both collisional ionisation and photoionisation.}
	    {The relative weakness of photoionisation does not stem from the intrinsic weakness of its AGN, whose X-ray luminosity is  $\approxgt 10^{42}$~erg~s$^{-1}$. We suggest that it may instead come from the line-of-sight from the active nucleus to the NLR being blocked by optically thick matter in the broad line region, at the same time responsible for the large observed variation of the column density obscuring the X-ray active nucleus. Alternatively, NGC~1365 could host a remarkably luminous nuclear starburst when compared to the AGN accretion power.}

\authorrunning{Guainazzi et al.}

\titlerunning{NGC~1365 soft X-ray emission}

\maketitle
%

\section{Ionising the NLRs: the X-ray view}

Most of the observational constraints on the
physics and geometry of narrow line regions (NLRs)
in active galactic nuclei (AGN) come from optical spectroscopy and
imaging. Comparison between spatially resolved spectroscopy of nearby
bright obscured AGN and photoionisation models (\cite{ferland86,osterbrock89})
and the study of ionisation cones (\cite{pogge88,tadhunter89})
indicate that the AGN high-energy emission is the main source
of ionising photons, although the contribution by collisionally
ionised plasma also seems to be required to accurately match the spectra
(\cite{viegasaldrovandi89}). A correlation between
radio power and line widths (\cite{wilson80,whittle85,whittle92,ulvestad01})
and the good morphological correlation between [OIII] and radio
images (\cite{capetti96,axon98}; see however \cite{das06} for
a different view) raise the possibility that the interaction between
radio jets and the NLR gas may affect the ionisation state of the NLRs
as well. Experimental support to this hypothesis
has so far been presented in one case only (\cite{axon98}).

In this context, X-rays have recently opened a complementary
window on NLR physics.
Whenever (extended) NLRs in nearby Seyfert~2
galaxies are larger than the
angular resolution of the {\it Chandra} optics ($\simeq$0.5$\arcsec$),
X-ray extended emission on scales from a few hundred to about 2~kpc
has been discovered (\cite{wilson00,young01,bianchi06}).
X-rays bear a remarkable morphological
similarity to images in the [[OIII] band.
Simple photoionisation models applied to the optical to
X-ray surface-brightness ratio indicates that solutions exist
in terms of a single phase AGN-photoionised medium, where the
ionisation parameter remains constant across the whole ENLR
(\cite{bianchi06}).
That implies a radial decrease of the density $\propto r^{-2}$.
This is generally consistent with results derived from optical spectroscopy
(\cite{kraemer00,bradley04,kraemer08}), although recent
high-resolution spatially-resolved spectroscopy in a small sample
of local Seyfert galaxies suggests a shallower dependence
(\cite{bennert06}), in agreement with older results on Mkn~573
(\cite{capetti96}).

High-resolution spectroscopy in the soft X-ray band (0.2--2~keV) confirms
the overall picture. In this energy band, spectra of obscured AGN
are typically dominated by recombination lines from He- and H-like
transitions from C to Si, and by a Fe-L transitions ``forest'', with negligible
continuum contribution (\cite{guainazzi07}; GB07 hereafter). In a few
bright sources, the quality of the spectrum warrants detailed spectral
diagnostics. In all these cases, photoionisation seems to be the dominant
ionisation mechanism (\cite{kinkhabwala02,sako00,sambruna01,armentrout07}).
In NGC~1068, by far the brightest obscured AGN of the soft X-ray sky,
the contribution from an optically thin collisionally ionised plasma
can be constrained to be $\approxlt 10\%$ of integrated soft X-ray flux
(\cite{brinkman02}).
Resonant scattering plays also an important role in the overall ionisation
balance (\cite{kinkhabwala02}); this constrains the gas column density
to $N_H \sim 10^{17 - 18}/Z_O$~cm$^{-2}$
(GB07), where $Z_O$ is the oxygen abundance.

In their spectroscopic study of a sample of 69 Seyfert~2 galaxies
observed with the XMM-Newton reflection grating system (RGS;
\cite{denherder01}),
GB07 suggest that the conclusions derived from detailed
spectral analysis of the brightest objects can be extended to the
underlying population of nearby Seyfert~2 galaxies up to a flux level
$\sim 10^{-13}$~erg~cm$^{-2}$~s$^{-1}$. However, GB07 noticed that
peculiar objects also exist, whose X-ray spectrum exhibits different
phenomenological properties from the rest of the sample.
NGC~1365
is the X-ray brightest of these ``oddities''.

NGC~1365 is an intriguingly complex galaxy.
Optically, it has been classified as a Seyfert either of
type 1.5 (\cite{veron80,hjelm96}), 1.8 (\cite{maiolino95}), or
2.0 (\cite{turner93}). Following GB07 and Bianchi et al.
(2009b), we consider in this paper NGC~1365 a ``type~2''
AGN on the basis of the column density covering its hard
X-ray emission ($> 1.5 \times 10^{23}$~cm$^{-2}$; \cite{risaliti07,risaliti09a}).
Its obscured active
nucleus is highly variable at different wavelengths
(\cite{veron80,hjelm96,risaliti07}). It hosts a circumnuclear
star-forming ring with a diameter of $\simeq$1.3~kpc
($\simeq$14$\arcsec$; 1$\arcsec$ $\simeq 90$~pc at the
distance of NGC~1365: $D = 18.6 \pm 1.9$~Mpc, \cite{silbermann99}).
This ring is resolved into many compact super star clusters
(\cite{kristen97}). Some of them are extremely massive and young
(\cite{galliano05,galliano08}). The latter were detected also
in the CO band (\cite{sakamoto07}), leading to an estimate
of molecular material mass of the order of $10^9 M_{\odot}$ within 2~kpc
from the AGN.
The nuclear starburst ring must be a copious source of X-rays, 
predominantly produced by two processes (\cite{mashesse08,grimm03,persic02}).
Softer thermal-plasma 
emission comes from gas heated to $\sim 10^6$~K by supernova-driven 
shocks propagating through denser interstellar medium. Harder, 
usually featureless X-ray emission comes from accreting compact 
objects (high-mass X-ray binaries).

At the same time, NGC~1365 hosts a
5$\arcsec$ bi-conical and asymmetrical
[{\sc O iii}] outflow (\cite{hjelm96,kristen97,veilleux03})
and a nuclear radio jet, the latter embedded in the cone and extending
5$\arcsec$ SE along the galaxy major axis (\cite{sandqvist95}).
The fact that the
[{\sc O iii}] $\lambda$5007 versus H$\alpha$ ratio in the cone is larger than
one suggests photoionisation by the AGN.
[{\sc N ii}]/H$_{\alpha}$ ratios $\ge$1 are measured out to 5--6 kpc 
from the nucleus (\cite{veilleux03}) and 
are consistent with either AGN photoionisation 
or shock ionisation.

For all the above reasons, NGC~1365 represents a unique laboratory
to study the connection between nuclear activity and star formation.
In this paper we primarily present and discuss
the second deepest high-resolution X-ray
spectrum of an obscured AGN ever (after NGC~1068).
It was collected
during an occultation experiment of the NGC~1365 AGN in 2007
(\cite{risaliti09b}). Archival data from an earlier (2004)
observation of the same field are also presented in this paper.

The paper is organised as follows: we describe the observations and
data reduction in Sect.~2; in Sect.~3 we present a phenomenological
analysis of the total RGS spectrum, aimed at
characterising the emission line content. In Sect.~4 we compare the
RGS spectrum with physical model of optically thin, collisionally
ionised plasma and photoionisation, and determine the contribution
of each mechanism to the emission line
spectrum. In Sect.~5 we compare our findings
with spatially resolved CCD-resolution spectroscopy obtained
with the {\it Chandra} ACIS, and discussed in
Wang et al. (2009). We put our results in the context of
the CIELO
({\it Catalogue of Ionised Emission Lines in Obscured AGN})
sample of Seyfert galaxies (GB07)
in Sect.~6, and summarise our main results in Sect.~7.
Unless otherwise specified, statistical uncertainties are
at the 90\% confidence level for one interesting parameter.

\section{Observations and data reduction}

The field around NGC~1365 has been observed several times by XMM-Newton.
In order to ensure an accurate reconstruction of the wavelength scale
in the RGS spectra, only observations aiming at the galaxy nucleus
(Tab.~\ref{tab7}) are considered in this paper.
\begin{table}
\caption{Log of the XMM-Newton observations of NGC~1365 presented
in this paper.}
\label{tab7}
\begin{center}
\begin{tabular}{lcc} \hline \hline
Observation \# & Start date & Exposure time \\
& & (ks, RGS1) \\ \hline
0205590301 & 17-Jan-2004 & 59.0 \\
0205590401 & 24-Jul-2004 & 67.8 \\
0505140201 & 30-Jun-2007 & 126.9 \\
0505140401 & 02-Jul-2007 & 127.1 \\
0505140501 & 04-Jul-2007 & 125.1 \\
\hline \hline
\end{tabular}
\end{center}
\end{table}

\begin{figure}
\hbox{
\includegraphics[height=6cm]{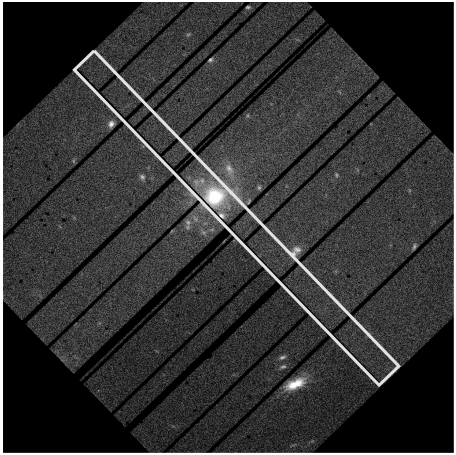}
}
\caption{RGS aperture ({\it box}) compared to the EPIC-pn
0.3--2~keV field of view (Obs.\#0505140201 in this
example)
}
\label{fig10}
\end{figure}

The RGS spectra were reduced following the guidelines in GB07.
We used the data reduction
pipeline {\tt rgsproc} in SASv7.1 (\cite{gabriel03}),
coupled with the most updated calibration files available in
August 2008. We choose a
fixed celestial reference point for the attitude solution coincident with
the NED optical nucleus of NGC~1365 ($\alpha_{2000} = 03^h33^m36.4s$,
$\delta_{2000} = -36^d08^m25s$).
The RGS aperture encompasses the whole NGC~1365 X-ray nucleus,
including the 5~kpc extended emission imaged by {\it Chandra}
(\cite{wang09}), as well as several point-like sources in the
galaxy (Fig.~\ref{fig10}).
Intervals of flaring particle background were filtered out by
applying a fixed threshold of 2 counts per second
on the background events, CCD\#9, $\Delta t = 10$~s
light curve.

Source spectra were extracted
in regions of the dispersion versus cross-dispersion and Pulse Invariant
versus cross-dispersion planes, corresponding to 95\% of the point
spread function (PSF) in the cross-dispersion direction. Background spectra
have been generated using a sub-set of blank field observations, whose
background count rate matches the level measured during each individual
RGS observation. This method is appropriate for sources with extended
emission on scales comparable to the PSF such as NGC~1365.

\begin{figure*}
\begin{center}
\includegraphics[height=17cm,angle=90]{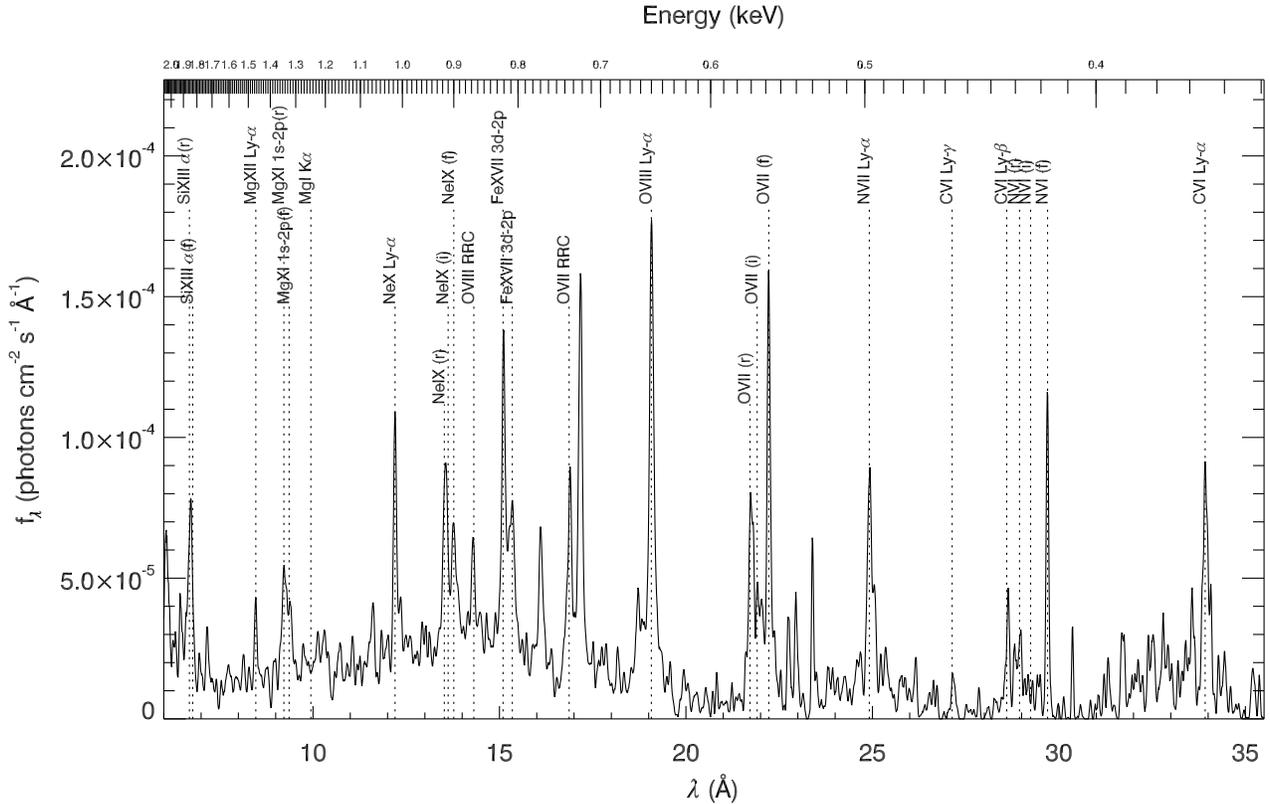}
\end{center}
\caption{
RGS net background-subtracted fluxed spectrum obtained
after merging together all spectra accumulated during the
XMM-Newton observations. A triangular smoothing with
a 5-channel kernel has been applied for visual purposes
only (in analogy to Fig.1 in GB07).
Atomic transitions observed in the RGS spectrum
of NGC~1365 are labelled for
reference.
}
\label{fig1}
\end{figure*}

The spectrum is dominated by emission
lines, as typically observed in obscured AGN (GB07),
although an continuum emission is also clearly visible, especially
in the wavelength range between 10 and 20\AA.
Such a continuum component cannot be associated to
transmission of the nuclear emission through a Compton-thin
state, because the column density in these states is always
higher than $1.5 \times 10^{23}$~cm$^{-2}$ (\cite{risaliti07,risaliti09a}).
This
component represents a first remarkable difference with respect to the
archetypal Seyfert~2 galaxy NGC~1068 (\cite{kinkhabwala02}).
We compared the intensities of the
eight brightest emission lines in the spectra accumulated during each
individual observation. In all cases, the intensities were
consistent within the statistical uncertainties.
We therefore accumulated all the RGS
data into a single spectrum (``RGS merged spectrum'' hereafter).
Its total exposure time is 501 and 439~ks in the RGS1 and RGS2, respectively.
In Fig.~\ref{fig1} we show the spectrum obtained merging all the
RGS observations of NGC~1365.

\section{The half Ms RGS spectrum: the line content}

As first analysis step, we have characterised the emission line
content of the spectrum. For this purpose, we fit segments of
100 spectral channels with a simple power-law continuum (spectral
index, $\Gamma$ fixed to 1; the quality of the fit is not
significantly improved if this parameter is left free in the fit)
corrected by photoelectric absorption due to neutral gas in
our Galaxy ($1.3 \times 10^{20}$~cm$^{-2}$
according to the LAB survey: \cite{kalberla05})
and as many emission lines as required according to the following
criterion: that the addition of a new emission line yields a
decrease in the value of the C-statistics (Cash) of 9.21, formally
corresponding to 99\% confidence level for two interesting
parameter (\cite{lampton76}). We fit He-like triplet components together,
imposing a difference among the component's centroid
energy as dictated by the atomic physics. Lines are
assumed in this paper to be unresolved, unless otherwise specified.

The results are summarised in Tab.~\ref{tab2}, where
we show the best-fit line energy and intensity, the intensity
normalised to the intensity of the forbidden component of
the He-$\alpha$ O{\sc vii} (``normalised intensity'' hereafter),
the normalised intensity in NGC~1068 according to our re-analysis
of the RGS data of this target, and the most likely
identification with the corresponding laboratory
energy. The errors in this Table are purely statistical
at the 1$\sigma$ level.
\begin{table*}
\caption{Energy and intensity of the lines detected in the combined RGS
spectrum of NGC~1365.}
\label{tab2}
\begin{center}
\begin{tabular}{c|c|c|c|c|c} \hline \hline
Energy (eV) & Intensity ($I$)& $I/I$[{\sc O vii}(f)] & $I/I$[{\sc O vii}(f)] & Transition ($E_{lab}$)$^a$ & Contribution by \\
(eV) & (10$^{-5}$~s$^{-1}$ cm$^{-2}$) & NGC~1365 & NGC~1068 & (eV) & thermal plasma (\%)$^b$ \\ \hline
$  367.4\pm0.2$ & $   1.4\pm^{  0.4}_{  0.5}$ & $   0.60\pm^{  0.18}_{  0.19}$ & $   1.16\pm^{  0.05}_{  0.19}$ &     C{\sc VI} Ly-$\alpha$ (367.56) & $100 \pm 50$ \\
$  430.72\pm^{ 0.02}_{ 0.09}$ & $   0.7\pm0.4$ & $   0.30\pm^{  0.15}_{  0.17}$ & $   0.35\pm^{  0.02}_{  0.17}$ &      N{\sc VI} He-$\alpha$ (r) (430.75)  & $90 \pm 50$ \\
(r) - 4.4 & $< 0.6$ & $<0.25$ & $0.08 \pm 0.02$ &      N{\sc VI} He-$\alpha$ (i) (426.37)  & $\le 100$ \\
(r) - 10.8 & $   1.8\pm0.3$ & $   0.73\pm^{  0.13}_{  0.14}$ & $   0.89\pm^{  0.03}_{  0.14}$ &      N{\sc VI} He-$\alpha$ (f) (419.86) & $26 \pm 7$ \\
$  435.26\pm^{ 0.16}_{ 0.30}$ & $   0.6\pm^{  0.2}_{  0.3}$ & $   0.27\pm^{  0.10}_{  0.11}$ & $  < 0.06$ &     C{\sc VI} Ly-$\beta$ (435.91) & $70 \pm 40$ \\
$  459.0\pm^{ 0.5}_{ 0.3}$ & $   0.40\pm^{  0.17}_{  0.18}$ & $   0.16\pm 0.07$ & $   0.11\pm^{  0.01}_{  0.07}$ &     C{\sc VI} Ly-$\gamma$ (459.42) & $90 \pm 50$ \\
$  499.96\pm^{ 0.04}_{ 0.22}$ & $   1.2\pm0.2$ & $   0.52\pm^{  0.09}_{  0.10}$ & $   0.41\pm^{  0.02}_{  0.10}$ &    N{\sc VII} Ly-$\alpha$ (500.31) & $100 \pm 30$  \\
$  573.99\pm^{ 0.05}_{ 0.05}$ & $   1.0 \pm 0.2$ & $   0.41\pm^{  0.09}_{  0.10}$ & $   0.38\pm^{  0.03}_{  0.10}$ &     O{\sc VII} He-$\alpha$ (r) (574.02) & $110 \pm 30$ \\
(r) - 5.3 & $   0.6\pm0.2$ & $   0.23\pm^{  0.09}_{  0.10}$ & $   0.12\pm^{  0.02}_{  0.10}$ &    O{\sc VII} He-$\alpha$ (i) (569.07) & $34 \pm 12$ \\
(r) - 13.0 & $   2.4\pm0.3$ & $   1.00\pm0.13$ & $   1.00\pm^{  0.04}_{  0.13}$ &   O{\sc VII} He-$\alpha$ (f) (561.06) & $40 \pm 8$  \\
$  653.4\pm0.6$ & $   2.5\pm0.2$ & $   1.03\pm^{  0.08}_{  0.09}$ & $   0.52\pm^{  0.02}_{  0.09}$ &   O{\sc VIII} Ly-$\alpha$ (653.70) & $110 \pm 20$ \\
$  726.00\pm^{ 0.35}_{ 0.13}$ & $   1.89\pm^{  0.18}_{  0.19}$ & $   0.78\pm0.08$ & $   0.14\pm^{  0.01}_{  0.08}$ & Fe{\sc XVII} 3$^s$2$^p$ (726.29)  & $86 \pm 11$ \\
$ 738.1\pm^{0.3}_{0.9}$ & $0.47 \pm^{0.15}_{0.13}$ & $0.20 \pm 0.06$ & $0.104 \pm 0.014$ & O{\sc vii} RRC$^c$ (739.11) & $60 \pm 20$ \\
& & & &  &  \\
$  773.54\pm^{ 0.02}_{ 0.34}$ & $   0.75\pm^{  0.13}_{  0.13}$ & $   0.31\pm^{  0.05}_{  0.06}$ & $   0.09\pm^{  0.01}_{  0.06}$ & O{\sc viii} Ly-$\beta$ (774.71)$^d$  & $83 \pm 16$  \\
$  824.57\pm0.07$ & $   1.49\pm^{  0.16}_{  0.17}$ & $   0.62\pm 0.07$ & $   0.12\pm^{  0.01}_{  0.07}$ & Fe{\sc XVIII} 3$^d$2$^p$ (824.72)  & $98 \pm 14$ \\
$  872.0\pm0.7$ & $   0.42\pm^{  0.12}_{  0.13}$ & $   0.18\pm 0.05$ & $   0.06\pm^{  0.01}_{  0.05}$ & OVIII RRC (871.52)$^e$  & $60 \pm^{70}_{40}$ \\
$  921.63\pm^{ 0.25}_{ 0.12}$ & $   0.42\pm^{  0.17}_{  0.18}$ & $   0.18\pm^{  0.07}_{  0.08}$ & $   0.10\pm^{  0.01}_{  0.08}$ &     Ne{\sc IX} He-$\alpha$ (r) (922.14)  & $130 \pm 60$\\
(r) - 7.2 & $   <0.33$ & $   <0.14$ & $   <0.05$ &     Ne{\sc IX} He-$\alpha$ (i) (915.13) & $< 100$ \\
(r) - 13.0 & $   0.40\pm^{  0.11}_{  0.12}$ & $   0.17\pm 0.05$ & $   0.10\pm^{  0.01}_{  0.05}$ &     Ne{\sc IX}He-$\alpha$ (f) (905.11) & $90 \pm 30$ \\
$ 1021.13\pm^{ 1.20}_{ 0.10}$ & $   1.0\pm0.2$ & $   0.43\pm^{  0.09}_{  0.10}$ & $   0.11\pm^{  0.01}_{  0.10}$ &     Ne{\sc X} Ly-$\alpha$ (1021.92) & $120 \pm 30$ \\
$ 1353.2\pm^{ 2.1}_{ 1.4}$ & $   0.57\pm^{  0.15}_{  0.16}$ & $   0.24\pm^{  0.06}_{  0.07}$ & $   <0.05$ &     Mg{\sc XI} He-$\alpha$ (r) (1352.24) & $110 \pm 15$ \\
(r) - 8.0& $<0.6$ & $<0.25$ & $<0.04$ &     Mg{\sc XI} He-$\alpha$ (i)  (1343.44) & $\le 100$ \\
(r) - 21.0 & $   0.33\pm^{  0.15}_{  0.16}$ & $   0.14\pm^{  0.06}_{  0.07}$ & $   <0.04$ &     Mg{\sc XI} He-$\alpha$ (f)  (1331.90) & $90 \pm 50$ \\
$ 1473.0\pm^{ 3.7}_{ 1.7}$ & $   0.50\pm^{  0.18}_{  0.19}$ & $   0.21\pm^{  0.07}_{  0.08}$ & $   <0.04$ &   Mg{\sc XII} Ly-$\alpha$ (1472.51) & $90 \pm 30$ \\
$ 1896.9\pm0.7$ & $< 47$ & $<33$ & $   <0.33$ &   Si{\sc XIII} He-$\alpha$ (r) (1867.47) & $\le 100$ \\
(r) - 13.8  & $< 2.4$ & $<1.0$ & $   <0.05$ &   Si{\sc XIII} He-$\alpha$ (i) (1853.61) & $\le 100$ \\
(r) - 27.8  & $   1.9\pm^{  1.1}_{  0.7}$ & $   0.8\pm^{0.5}_{  0.3}$ & $   <0.04$ &   Si{\sc XIII} He-$\alpha$ (f) (1839.76) & $90 \pm 30$ \\
$ 1892.7\pm0.8$$^f$ & $   1.1\pm0.4$ & $   0.46\pm 0.18$ & $   <0.05$ & & ... \\
\hline \hline
\end{tabular}
\end{center}

\noindent
$^a$most likely identification. The laboratory energy is shown in brackets.
$^b$contribution to the total intensity due to the best-fit model in Tab.~3. Values higher than 100\% mean that the model
over-predicts the data at that energy;
$^c$fit with a single Gaussian ($\sigma \le 8$~eV). Possible blending with
Fe {\sc xvii} 2$^s$2$^p$ (738.98~eV): see Sect.~4.1 for the
results of a fit with two Gaussians;
$^d$possible blending with Fe{\sc xviii} 3$^s$2$^p$ (773.02~eV);
$^e$fit with a single Gaussian ($\sigma \le 3$~eV). Possible blending with Fe{\sc xviii} 2$^s$2$^p$ (872.75~eV). See Sect.~4.1 for the
results of a fir with two Gaussians;
$^f$unidentified.
\end{table*}
Only lines detected according to the aforementioned statistical criterion
are shown, except for the He-$\alpha$ triplets, for which all
the components are shown, even if not all of them
are formally detected.

The spectrum of NGC~1365 is remarkably different from that of the
archetypal Seyfert~2 Galaxy NGC1068.
Once the line intensities are normalised
to the forbidden component of the O{\sc vii} He-$\alpha$ triplet, we find that:

\begin{enumerate}
\item Ly-$\alpha$ transitions of H-like species are
stronger in NGC~1365 (except C{\sc vi} Ly-$\alpha$)
\item Fe-L transitions are stronger in NGC~1365
\item the intensities of the He-like magnesium and
silicon triplet components are higher in NGC~1365,
although the poor statistics of these features
in the NGC~1068 spectrum prevents us from saying whether
there is a difference also in the relative weight of the
triplet components
\end{enumerate}

We'll discuss here briefly these outcomes,
because they provide us with a guideline on the
spectral analysis developed in Sect.~4 and
Sect.~5. A more detailed discussion of
the astrophysical implications of the
soft X-ray spectrum of NGC~1365 is, however,
deferred to Sect.~6.

The forbidden transitions of He-like triplets are strongly
suppressed in collisionally ionised plasmas. This
provides an explanation for Item\#1 above in terms
of a larger contribution of a collisionally ionised plasma
in NGC~1365.
Indeed, a high ratio between the intensity
of the Ly-$\alpha$ O{\sc viii} against the
forbidden component of the O{\sc vii} He-${\alpha}$ is
an empirical criterion to distinguish
starburst from AGN in high-resolution soft X-ray
spectra (GB07). A strong contribution by Fe-L
transition may also be expected in thermal plasmas
(\cite{mewe85}). This may provide an explanation
for Item\#2. Nonetheless, comparatively strong
iron lines could also be the signature of metal
overabundance. This could also explain the high
intensities of magnesium and silicon triplet components.
Wang et al. (2009) found regions of high Si- and
Fe-abundances N and NW the NGC~1365 nucleus
in their analysis of the deep high-resolution
{\it Chandra}/ACIS image (together with regions
of low Si-abundances primarily to the South the nucleus).
We will later discuss the issue of
heavy elements metallicity in the NGC~1365
interstellar medium.
On the other hand,
the intensities of the He-like nitrogen, oxygen and
neon He-$\alpha$ triplet components are comparable in
NGC~1068 and NGC~1365.

A feature at centroid energy $E \simeq 738.1$~eV
($\lambda = 16.800$~\AA)
is close to the O{\sc vii} Radiative Recombination
Continuum (RRC). This feature is
narrow. If
fit with a Gaussian profile the upper limit on its
intrinsic width is 0.8~eV.
Narrow RRC features are unambiguous signatures of
photoionised plasmas (\cite{liedahl95}).
Nonetheless, contamination by the nearby
Fe{\sc xvii} 2$^s$2$^p$ feature at
16.780\AA\, is possible.
We will
further discuss the identification of this feature
in the framework of the global thermal
fit of the RGS spectrum (Sect.~4.1).

\section{Spectral fitting of the RGS spectrum}

The next step for the understanding of the physical processes responsible
for the soft X-ray emission in NGC~1365
is to use physical emission models to fit globally the whole soft
spectrum or, at least, to simultaneously fit
the emission lines in Tab.~\ref{tab2}.
We have followed two complementary approaches: global fit of the 
RGS spectrum with a combination of thermal components;
and simultaneous fits of emission line ratios through
CLOUDY-based photoionisation models.

\subsection{Thermal fit of the overall RGS spectrum}

Firstly we have tried to fit the combined RGS spectra with a combination
of optically thin, collisionally ionised plasma
(``thermal scenario'' hereafter).
We used trough this paper the {\tt apec} model (\cite{smith01}) in
{\sc Xspec v12} (\cite{arnaud96}) to model such a component.
We used as a criterion to increase the model complexity that any
further component
yields an improvement of the C-statistic $\Delta C \le -9.21$.
Following the
results from
the spatially-resolved spectroscopy of the {\it Chandra} diffuse emission
(\cite{wang09}), we have assumed that each continuum spectral component
is modified by its own photoelectric absorption column density.
In all the models we also included one power-law and one blackbody
continuum. These components are required to fit emission excesses
in the softest ($E \le 0.4$~keV) and hardest ($E \ge 1.7$~keV)
RGS band.
The blackbody component may
account for the integrated emission of point sources
(X-ray binaries and Ultra Luminous X-ray sources, ULX).
Alternative parametrisation
for this ``soft excess'' are possible, such as emission
from a disk blackbody or bremsstrahlung. They do not
substantially affect the main results present in this paper.
The power-law component is instead required to fit a featureless
hard excess above $\simeq 1.7$~keV. Only an upper limit
on the spectral index is obtained, indicating
a very hard spectrum ($\Gamma < -2.5$). Other parametrisation
of this component are also possible: a bremsstrahlung
yields, for instance, $kT > 28$~keV. Although in principle this spectral
component could be associated with the integrated emission
of unresolved X-ray binaries in the RGS aperture, it is
difficult to infer any meaningful physical constraints
given the small energy range on which it yields a significant
contribution to the fit. We will therefore refrain from
discussing it any further in this paper. We only observe
that the same component with a similar very flat spectrum
is required by the fit of the
NGC~1365 ACIS spectrum, by contrast to the soft continuum
component which is required by the RGS data only (cf. Sect.~5). 
Finally, we included
a photoelectric absorption component covering the whole RGS model, to
take into account obscuration by gas in the Galaxy along the line-of-sight
to NGC~1365.

\begin{figure}
\begin{center}
\includegraphics[height=8cm,angle=-90]{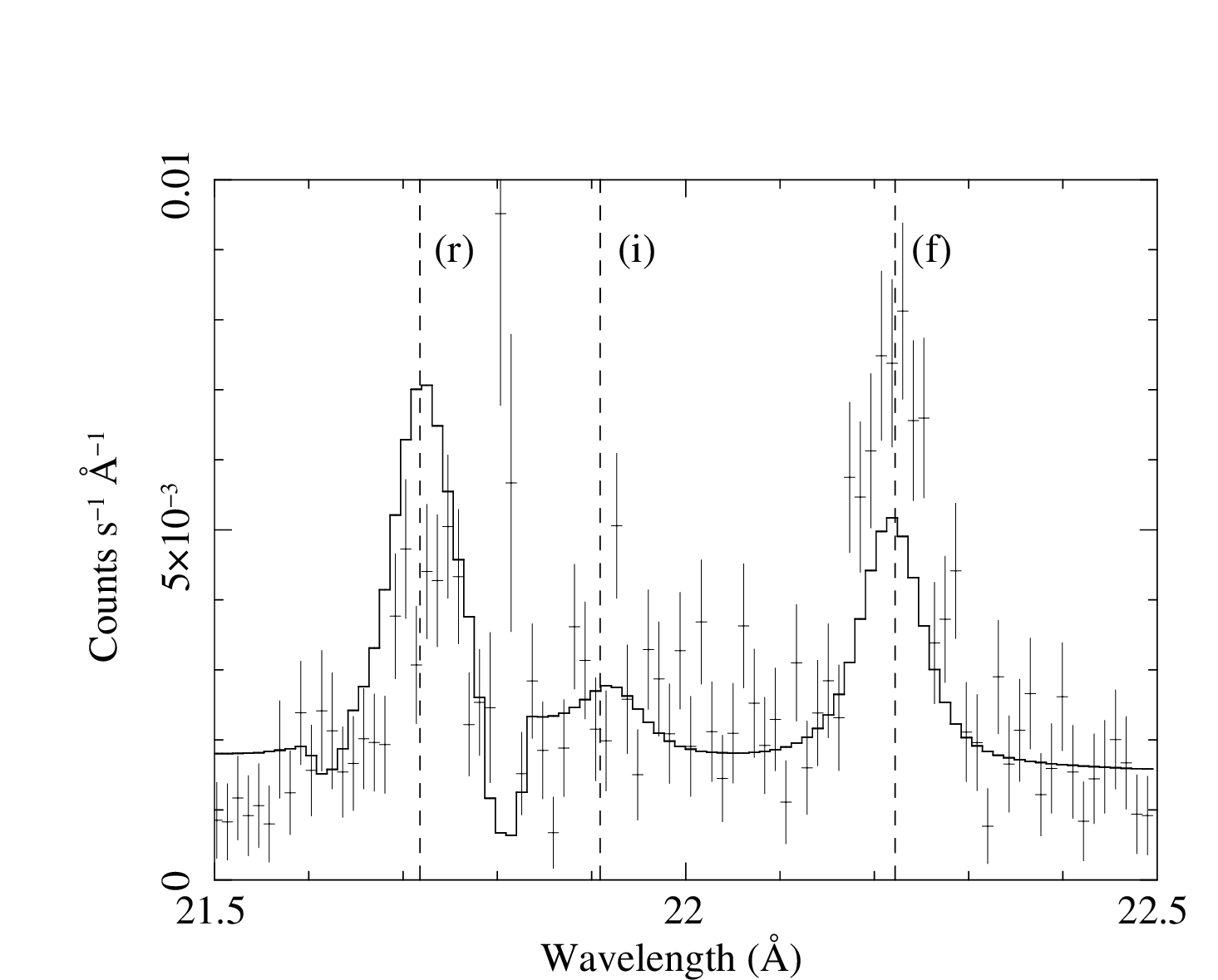}
\end{center}
\caption{
RGS spectrum around the O{\sc vii} He-$\alpha$ triplet
({\it crosses}) and best-fit model ({\it lines})
in a scenario where only optically thin collisionally ionised
plasma components (three in total) are used.
The observer's frame wavelength of the
forbidden (f), intercombination (i) and resonant (r)
components is marked by the {\it dashed lines}.
The apparent absorption feature at $\lambda \simeq$21.8\AA
is due to a bad detector pixel.
}
\label{fig2}
\end{figure}

The model outlined above, although capable of
reproducing most of the emission line intensities, fails at
reproducing the
O{\sc vii} and N{\sc vii} triplet ratios. This is illustrated in Fig.~\ref{fig2}
which shows the spectral region around the former feature superposed with
a model with three thermal plasma components. As expected for the
temperatures (in the range 0.2--1~keV) which these fits yield, the
model has a much higher recombination component than
required by the data.
Although it is impossible to associate an absolute confidence level
of the fit quality when the C-statistics is used, its high
value (13042.7/5304~degrees of freedom) is indicative of an imperfect model.

We have therefore removed from the fit wavelength ranges around
the O{\sc vii} (21.0-23.0\AA) and the N{\sc vi} (28.0-30.0 \AA) triplets,
and repeated the fit procedure. As a cross check we have verified
that the best-fit model obtained on this restricted wavelength range
does not overproduce the observed counts in the excised wavelength ranges.

\begin{figure*}
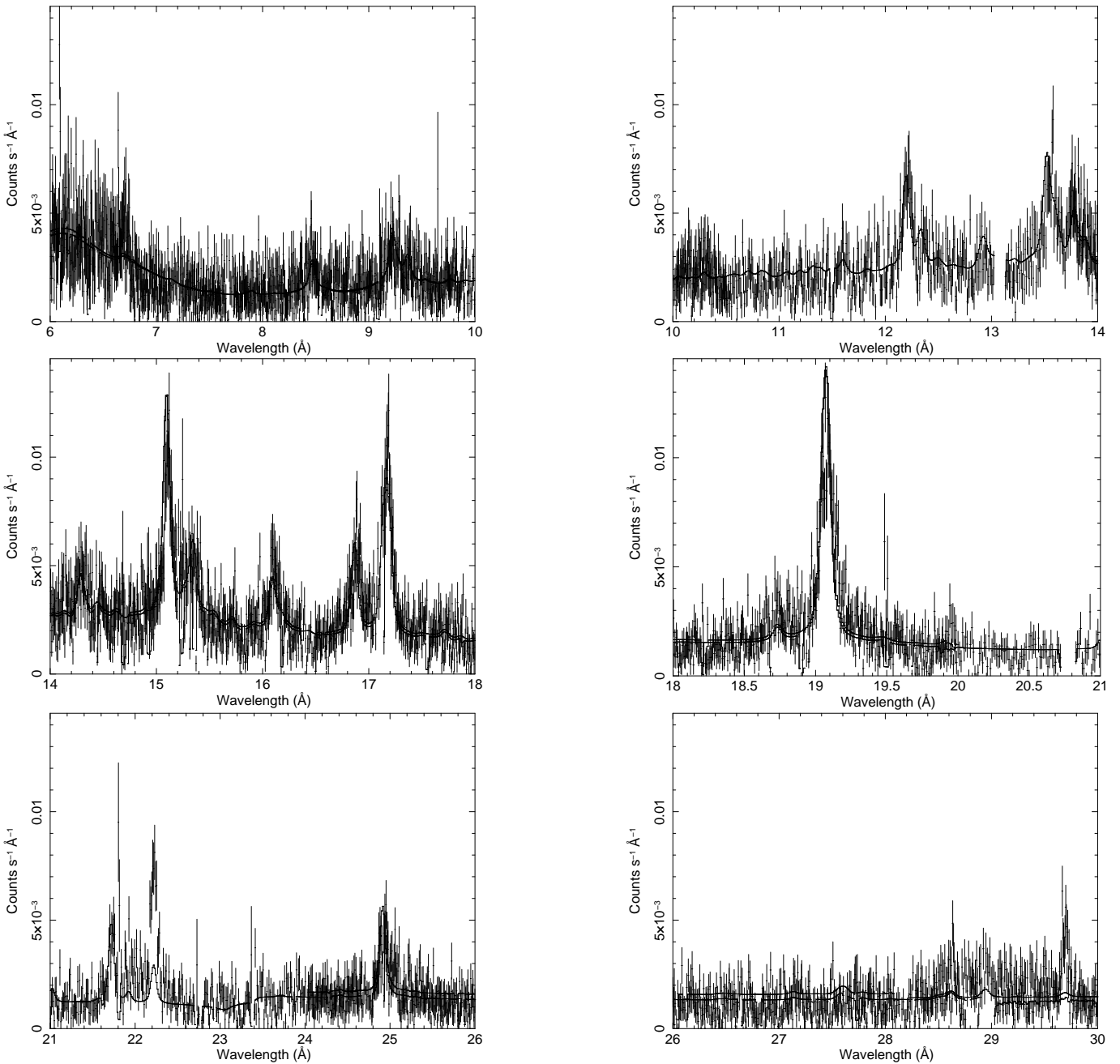

\begin{center}
\hbox{
\includegraphics[height=8cm,angle=-90]{fig3a.ps}
\hspace{2.0cm}
\includegraphics[height=8cm,angle=-90]{fig3b.ps}
}
\hbox{
\includegraphics[height=8cm,angle=-90]{fig3c.ps}
\hspace{2.0cm}
\includegraphics[height=8cm,angle=-90]{fig3d.ps}
}
\hbox{
\includegraphics[height=8cm,angle=-90]{fig3e.ps}
\hspace{2.0cm}
\includegraphics[height=8cm,angle=-90]{fig3f.ps}
}
\end{center}
\caption{
RGS spectrum ({\it crosses}) and best-fit model ({\it lines})
for the whole RGS sensitive bandpass when the best-fit
thermal model is applied. The wavelength scale is in the
observer's frame.
}
\label{fig3}
\end{figure*}

Using the restricted wavelength range, the best-fit requires only
two thermal components. In principle the abundances of each component
with respect to the solar values (\cite{anders89})
have been left free to vary independently. However, for only a sub-set
of them the corresponding confidence interval are significantly constrained.
They are listed in Tab.~\ref{tab3} together
\begin{table}
\caption{{\it Left column}:
Parameters for the RGS and {\it Chandra}/ACIS
thermal model best fits.}
\label{tab3}
\begin{center}
\begin{tabular}{lcc} \hline \hline
Parameter & RGS & ACIS \\ \hline
\multicolumn{3}{l}{Thermal components} \\
$N_{H,1}$ (10$^{20}$~cm$^{-2}$) & $< 0.4$ & $< 1.2$ \\
$kT_1$ (eV) & $301 \pm 7$ & $400 \pm 130$ \\
$N_1$$^a$ & $6.9 \pm ^{1.0}_{0.6}$ & $1.3 \pm^{1.5}_{1.1}$ \\
$N_{H,2}$ (10$^{20}$~cm$^{-2}$) & $34 \pm^3_2$ & $< 100$ \\
$kT_2$ (eV) & $690 \pm^{20}_{30}$ & $650 \pm^{120}_{50}$ \\
$N_2$$^a$ & $8.6 \pm^{0.3}_{0.9}$ & $1.8 \pm^{0.3}_{0.7}$ \\ \hline
\multicolumn{3}{l}{Elemental abundances with respect to solar} \\
$Z_C$  & $0.031 \pm^{0.012}_{0.010} $ & $< 1.08$ \\
$Z_N$  & $0.072 \pm^{0.017}_{0.011} $ & $< 0.15$ \\
$Z_O$  & $0.0123 \pm^{0.0022}_{0.0013} $ & $0.031 \pm^{0.030}_{0.015}$ \\
$Z_{Ne}$  & $0.030 \pm^{0.003}_{0.004} $ & $0.11 \pm^{0.03}_{0.04}$ \\
$Z_{Mg}$  & $0.083 \pm^{0.012}_{0.011} $ & $0.14 \pm 0.03$ \\
$Z_{Si}$  & $0.083 \pm^{0.012}_{0.011} $ & $0.15 \pm^{0.05}_{0.06}$ \\
$Z_{Fe}$  & $0.023 \pm 0.002 $ & $0.043 \pm ^{0.011}_{0.005}$ \\ \hline
\multicolumn{3}{l}{Blackbody} \\
$kT$ (eV) & $364 \pm 2$ & ... \\
$N$$^b$ & $3.1 \pm^{0.8}_{0.2}$ & ... \\
\hline \hline
\end{tabular}
\end{center}

\noindent
$^a$in units of $\frac{10^{-17}}{4 \pi [D_A (1+z)^2]} \int n_e n_H dV$,
where $D_A$ is the angular size of the source and $n_e$ and $n_H$ are
the electron and H densities, respectively

\noindent
$^b$in units of 10$^{-4} \frac{L_{39}}{D^2_{10}}$, where
$L_{39}$ is the source luminosity in units of 10$^{39}$~erg~s$^{-1}$
and $D_{10}$ is the distance to the source in units of 10~kpc

\end{table}
with the other best-fit parameters.
The measured Galactic
column density ($N_{H,Gal} = (8 \pm^3_5) \times 10^{19}$~cm$^{-2}$)
is marginally consistent with the contribution estimated
from the radio LAB survey. The C-statistic
value is 6237.0/4440~degrees of freedom.
In Fig.~\ref{fig3} we show the RGS spectrum and
best-fit thermal model as in Tab.~\ref{tab3} superposed.
The contribution to the total intensity of each emission line
due to this model (``baseline model'' hereafter)
is given in the rightmost column of
Tab.~\ref{tab2}.

A multiple-temperature structure could be the results of
fitting a single temperature non-equilibrium ionisation
plasma. We have tested this hypothesis by alternatively
fitting the RGS spectrum with a single {\tt vnei}
(\cite{borkowski01}) model
 in {\sc Xspec}. The resulting C-statistics is worse
(6857.7/4422~dof). Again, this in itself is not enough
to rule out this model. Nonetheless, we observe that
several lines in the short wavelength regime ($\lambda
\approxlt 14$~\AA) are not well accounted for as in
the baseline model with two equilibrium thermal components.
Moreover, it remains true also with the non-equilibrium
model that most of the forbidden component of the
O{\sc vii} and N{\sc vi} triplets are not accounted for.
The discrepancy in the case of the non-equilibrium collisional
scenario is even higher than in our baseline model.
The recombination (forbidden) component of the
O{\sc vii} He-$\alpha$ in the model is twice as strong
(four times weaker) than in the data. We will not consider
this solution any longer.

We discuss now the identification of
the feature at
$E = 738.1 \pm^{0.9}_{0.3}$~eV
($\lambda= 16.799 \pm^{0.020}_{0.007}$ \AA).
Formally, its wavelength is inconsistent with the
strongest close transitions: O{\sc vii} RRC
and Fe {\sc xvii} 2$^s$2$^p$ (at 
$\lambda$=16.777 and 16.780 \AA, respectively).
$60 \pm 20$\% (1-$\sigma$ error)
of its intensity are accounted by the thermal scenario
best-fit model, although some excess at lower
energies remain. We have tried to fit this excess
with an additional Gaussian emission line,
keeping its centroid energy frozen to the
source frame laboratory wavelength of
the O{\sc vii} RRC. The
fits improves by $\Delta C/\Delta \nu = 52.3/2$~dof.
The width and intensity of this additional component are:
$\sigma_{OVII RRC} = 14 \pm^7_9$~eV, and $I_{OVII RRC} =
(9\pm 4) \times 10^{-6}$~photons~cm$^{-2}$~s$^{-1}$.
We will refer to these values hereafter when discussing the
properties of the O{\sc vii} RRC in NGC~1365.

Similarly, the identification of the emission line
at $E \simeq 872$~eV ($\lambda = 14.218 \pm 0.011$~\AA)
is not fully unambiguous. The centroid energy is consistent
with either O{\sc viii} RRC ($\lambda =$14.228~\AA)
of Fe{\sc xviii} 2$^s$2$^p$ ($\lambda =$14.220~\AA).
A fit with two Gaussians having centroid energy fixed to
the laboratory values, however, yields only an upper limit
($\le 4 \times 10^{-6}$~cm$^{-2}$~s$^{-1}$) on the
intensity of the latter. With this modelling
the best-fit parameters of the O{\sc viii} RRC
are: $E = 870 \pm 2$~eV; $\sigma \le 8$~eV;
$I = (1.8 \pm^{2.0}_{1.3}) \times 10^{-6}$~cm$^{-2}$~s$^{-1}$.

We have also tried to constrain the bulk
velocity of the lines, by allowing the redshift
of the thermal components to vary. Indeed,
this yields an improvement in the quality of
the fit by $\Delta C/\Delta \nu = 121.3/1$~dof
for a (formal) velocity shift of
$\Delta v = 181 \pm 7$~km~s$^{-1}$.
This is comparable to the
systematic uncertainties in the RGS wavelength scale
(20~m\AA\, at 3$\sigma$, corresponding to
$\simeq$300~km~s$^{-1}$
at the wavelength of the O{\sc vii}
He-$\alpha$). We consider therefore
this piece of evidence
inconclusive as to whether the soft
X-ray lines in NGC~1365 are affected by
a velocity shift of astrophysical nature.

Assuming the baseline model as in Tab.~\ref{tab3},
the observed 0.3-2.0~keV flux is
$(2.61 \pm 0.03) \times 10^{-12}$~erg~cm$^{-2}$~s$^{-1}$,
corresponding to a luminosity (corrected for
Galactic absorption) of $1.73 \times 10^{41}$~erg~s$^{-1}$.

\subsection{CLOUDY-simulations of the line emission intensity ratios}

In order to estimate the contribution of photoionisation to the
soft X-ray lines, we have generated a grid of models of a photoionised
nebula with {\sc Cloudy} (\cite{ferland98}). We have assumed a
standard AGN continuum, with the following parameter: ``blue bump''
temperature $10^6$~K; $\alpha_{ox} = -1.5$; $\alpha_{UV} = -0.5$;
and $\alpha_{x} = -1.35$ (\cite{risaliti00}),
where the $\alpha$ are spectral energy
index in the band or between the bands specified by the subscripts.
$\alpha_{ox}$ was determined from the strictly simultaneous
optical-to-X-ray Spectral Energy Distribution measured by
the XMM-Newton EPIC and OM instruments during Compton-thin
states (see Risaliti et al. 2009a, 2009b for an analysis
of the EPIC spectra).
We have generated grid of reflected spectra from a photoionised nebula,
spanning a large range  in ionisation parameter:
$\log(U)$=[-2.0,2.0]\footnote{the ionisation
parameter $U$ is defined as $\frac{\Phi(H)}{n(H) c}$,
where $\Phi(H)$ is the surface flux of ionising photons,
and $n(H)$ is the total hydrogen density}; electronic density $\log(n_e)$=[3,14] and
total column density $\log(N_H)$=[17,24] in steps of 0.1 dex.
We have extracted intensity ratios from
these simulations for all the lines detected
in CIELO (GB07). For each model of the grid we
have calculated a figure of merit of the agreement between
the observed and the simulated line ratios through a generalised
$\chi^2$ function:
$$
\chi^2 = \sum \frac{(I_C-I_o)^2}{\sigma_{I_o}^2}
$$
where $I_o$ is the observed line intensity (with its
statistical error $\sigma_{I_o}$) and $I_C$
is the intensity predicted by {\sc Cloudy}, both normalised
to the value of the forbidden component of O{\sc vii} He-${\alpha}$.
The sum is carried out on all the line species
detected in the RGS spectrum of NGC~1365.
We consider acceptable all models, which
yield a $\chi^2$ value $\chi^2 \le \chi^2_{min}$~+9.21.

No solution exists, which can simultaneously fit the whole line
spectrum observed in NGC~1365 in terms of one photoionised
component, even if allowance is made for the elemental
abundances to vary in the range [0,10 $Z_{\odot}$].
Even more constraining, no solution
is able to simultaneously fit the intensity of the O{\sc viii}
Ly-${\alpha}$ and of the recombination component of the
O{\sc vii} He-${\alpha}$ triplet.
Fits with a combination of two
pure photoionisation models yields still
$\chi^2_{\nu} > 5$.
On the other hand, the $\chi^2$ value corresponding to the
fit with the RGS baseline model ({\it i.e.} two thermal
components without any
photoionisation components) is 47.0 for 19~dof, significantly
better than for two photoionisation components, although
still unacceptably high. A fit with one thermal
and one photoionisation component is only
marginally better: $\chi^2$=41.5 for 19~d.o.f.
(for $\log U \simeq$1.4).

These results, coupled with the spatially resolved spectroscopy 
of the {\it Chandra} NGC~1365 field (\cite{wang09}) leads
us to investigate in this section a physical scenario,
where the baseline thermal model developed in Sect.~4.1 applies,
and any photoionised component is required only to explain the
deviations between the experimental data and this baseline model.

\begin{figure}
\hspace{-1.0cm}
\includegraphics[height=8cm]{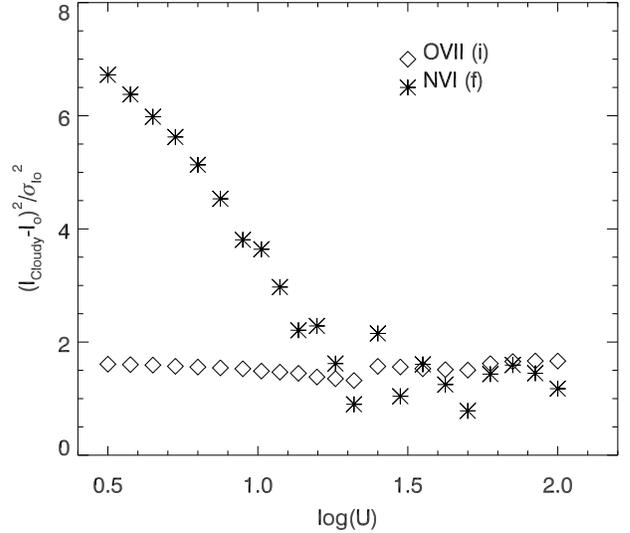}
\caption{$\chi^2$ contribution for
the N{\sc vi} (f) and O{\sc vii} (i)
lines as a function of $\log(U)$.
}
\label{fig5}
\end{figure}

\begin{figure}
\hspace{-1.0cm}
\vbox{
\includegraphics[height=8cm]{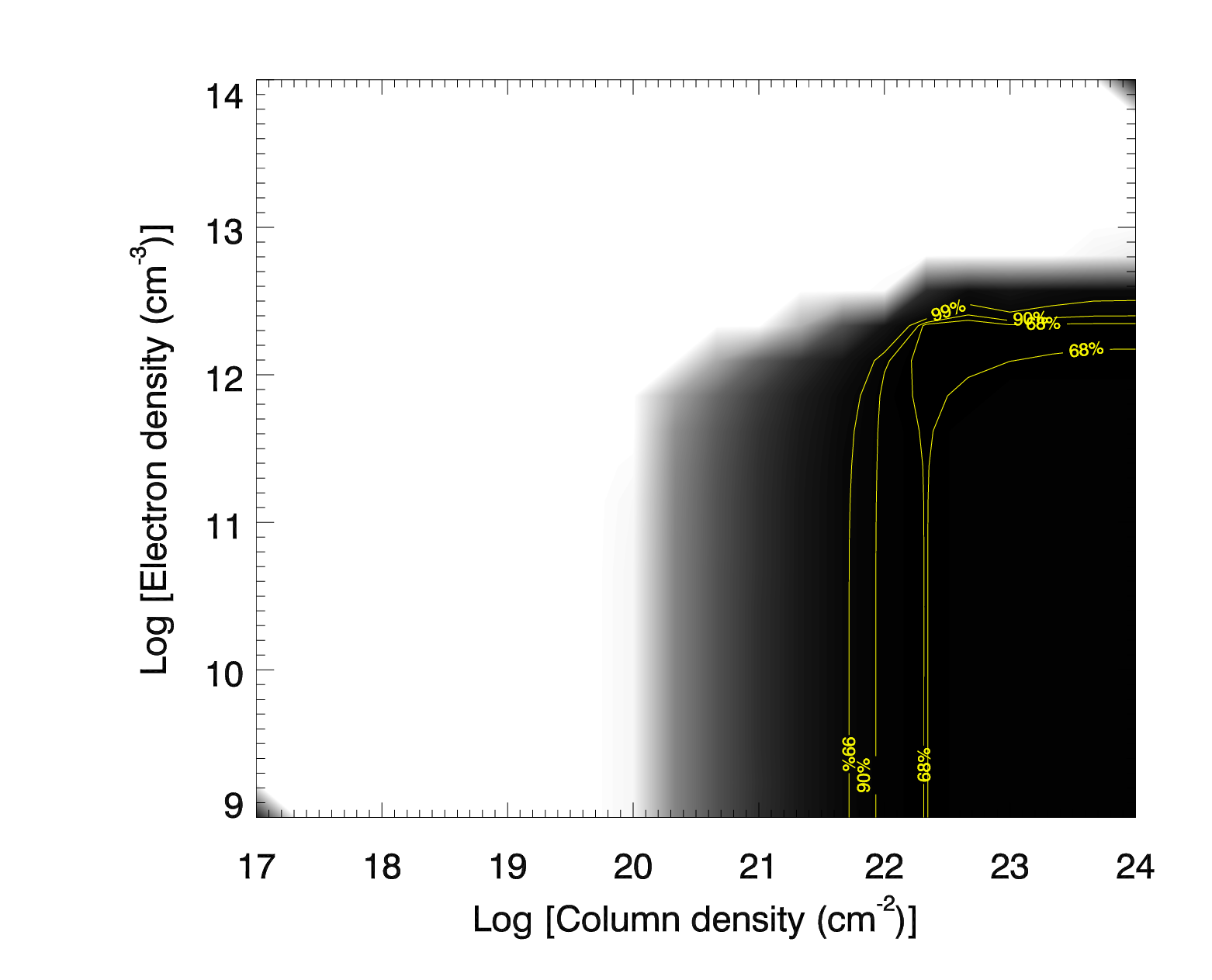}
\includegraphics[height=8cm]{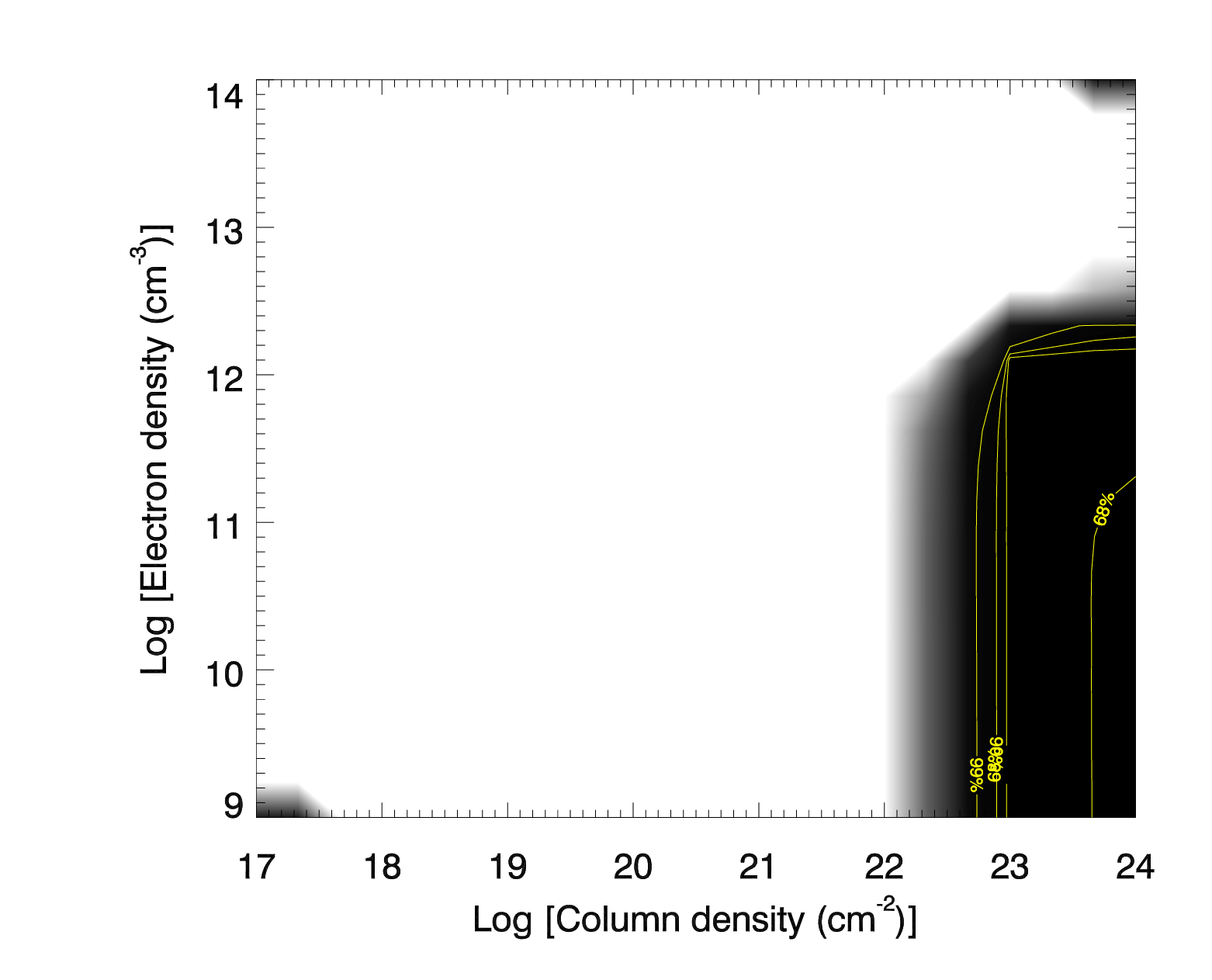}
\includegraphics[height=8cm]{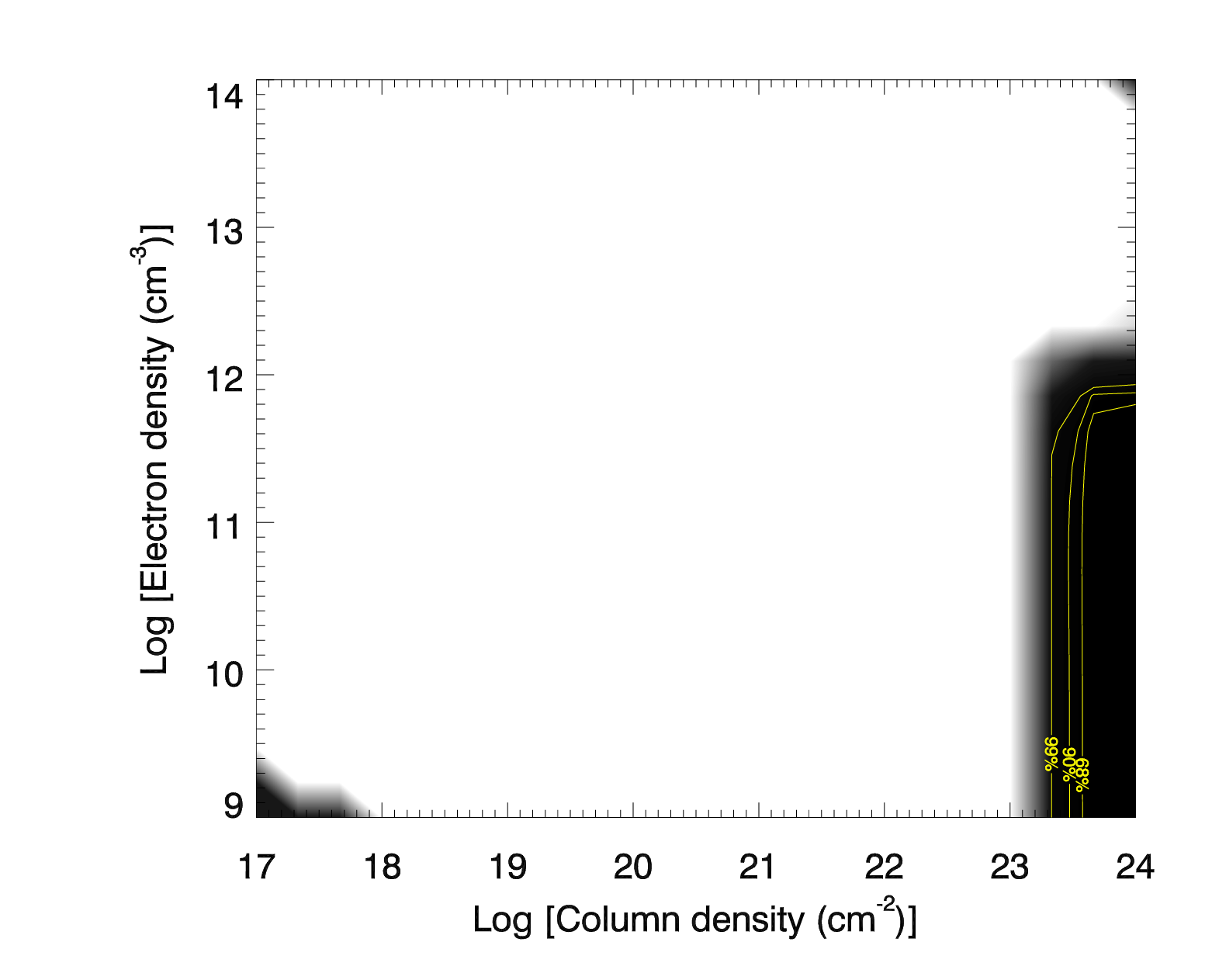}
}
\caption{Iso-$\chi^2$ contour plots in the
$n_e$ versus $N_H$ plane for $\log(U)$=1.2
({\it upper panel}), $\log(U)$=1.625
({\it middle panel}), and $\log(U)=1.925$
({\it lower panel}) The lines indicate
the confidence level at 68\%, 90\% and
99\% for two interesting parameters.
The contours are smoothed with a 3$\times$3
cells median filter for plotting purposes only. 
}
\label{fig4}
\end{figure}

We have therefore applied the grid of {\sc Cloudy} models to
a line spectrum obtained by the RGS observation after subtracting the
contribution of the thermal scenario best-fit according
to the rightmost column in Tab.~\ref{tab2}.
Only four of the lines detected in the RGS spectra have still intensity
measurements statistically different from zero
after the ``thermal'' contribution
is subtracted: the forbidden and inter-combination component of
O{\sc vii} He-$\alpha$,
the forbidden components of N{\sc vi} He-$\alpha$, and
(marginally) the O{\sc viii} Ly-$\beta$
transition,
as well as the O{\sc vii} and O{\sc viii} RRC, which, however,
are not simulated
by {\sc Cloudy}.
These four lines obviously
yield only very loose constraints on the physical properties
of the photoionised plasma responsible for their emission.
The minimum $\chi^2$ value is
20.3 for 20~dof, corresponding to $\log(U) = 1.625$. 
In principle values of $\log(U)$ comprised between 0.6 and 1.9
yield all $\chi^2$ value in the acceptance range.
One may use the contribution to the $\chi^2$ of the
N{\sc vi} forbidden line (Fig.~\ref{fig5})
to further constraint, at a lower statistical significance,
the range of acceptable ionisation parameters. At
1$\sigma$: $\log(U) \simeq 1.6 \pm^{0.3}_{0.4}$ (Fig.~\ref{fig5}).
In Fig.~\ref{fig4} we show iso-$\chi^2$ contour plots in the
$n_e$ versus $N_H$ plane for $\log(U)=$1.2,
1.625, and 1.925. The electron density is only loosely constrained
to be $\approxlt 10^{10}$cm$^{-3}$ and the column
density $\approxgt 10^{22}$cm$^{-2}$ at the
99\% confidence level for two interesting parameters
(these intervals are only marginally
more tightly constrained if
a less demanding confidence interval is chosen).

We stress that these results need not to be
over-interpreted, given the simplicity of the model assumed
and the strong assumptions underlying it. Our results are to
be interpreted more as a
simple ``proof of existence'' of photoionised
material in the nuclear region of NGC~1365 rather than as an
accurate measure of the photoionised spectrum.

\section{Comparison with the {\it Chandra} results}

Deep {\it Chandra} observations of NGC~1365 (\cite{wang09})
have shown that the
soft X-ray morphology is rather complex, with four components
contributing to the overall energy output: a) the unresolved
nucleus; b) diffuse emission on scale as large as 5~kpc; c)
off-nucleus point sources; d) the jet. Comparing the RGS spectroscopic
measurement with CCD resolution spatially-resolved spectroscopy
is therefore far from trivial, given the large RGS
aperture (cf. Fig.~\ref{fig10}). Only a fraction of this
field is imaged by {\it Chandra}.
Nonetheless, a comparison with the {\it Chandra}
spatially-resolved spectroscopic
results may be interesting to evaluate the systematic uncertainties
induced by spatially integrated spectroscopy on such a complex source.
We remind that the Point Spread Function of the
XMM-Newton optics is too large for the diffuse
soft X-ray emission associated to the circumnuclear starburst
to be discernible.

\begin{figure}
\includegraphics[height=9cm, angle=-90]{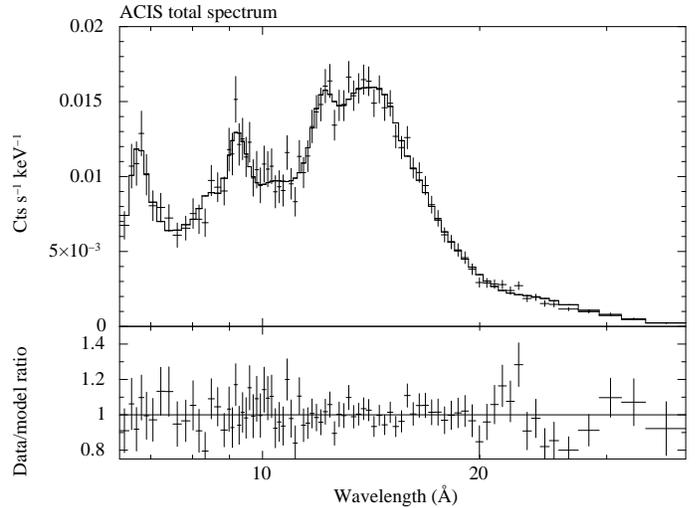}
\caption{ACIS total NGC~1365 spectrum
({\it upper panel}) and residuals against
the best-fit RGS model ({\it lower panel};
best-fit parameters in the rightmost column
of Tab.~\ref{tab3})
}
\label{fig6}
\end{figure}

We have extracted a {\it Chandra}-ACIS
spectrum from the whole NGC~1365 X-ray surface,
including nucleus and point sources (details on the data analysis
reduction can be found in Wang et al. 2009), and applied the
best-fit RGS baseline
model in the 0.3-2~keV energy band. The model provides a good
fit to the data ($\chi^2 = 62.7$/70~dof), and residuals are visible
mainly at an energy consistent with the O{\sc vii} He-$\alpha$ triplet
(Fig.~\ref{fig6}),
as expected.
The cross-calibration between RGS and the
{\it Chandra} ACIS is discussed by Plucinsky et al. (2008).
Flux measurements at the O{\sc vii} energies by
ACIS are typically lower (by $\simeq$10\%)
than the RGS. The difference becomes
smaller
($\pm 5\%$) at the Ne{\sc ix} and Ne{\sc x} energies.
The best-fit model parameters are shown in the
rightmost column of Tab.~\ref{tab3}. Interestingly enough,
the {\it Chandra} fit does not require the blackbody component.
Taking into account the
larger statistical uncertainties associated with the determination
of the parameters from the {\it Chandra} spectrum, the main
differences to the RGS best-fit are the normalisations of the
collisionally ionised plasma components and the metallicities.
The best-fit metallicities measured by {\it
Chandra} are systematically higher by a factor 2 to 4
than the RGS one (although consistent within
1$\sigma$). On the other hand, the normalisations of the continua 
measured by the ACIS are around a factor of 5 lower than measured
by RGS. We interpret these findings as due to the contribution of unresolved
point sources in the RGS aperture, which dilute the contrast between
the emission lines and their underlying continuum.
Indeed, point sources detected by {\it Chandra}
(among them two ULXs, \cite{strateva08,soria09}) contribute
around two thirds of the overall integrated soft X-ray flux.
In principle, one could sum up the spectra of these sources,
and subtract their contribution to the RGS spectrum. However,
given the fact that the XMM-Newton and {\it Chandra} observations
are not simultaneous, and that these sources are highly
variable (\cite{soria09}), the outcome of this exercise would
be merely academic.

\begin{figure}
\includegraphics[height=9cm, angle=-90]{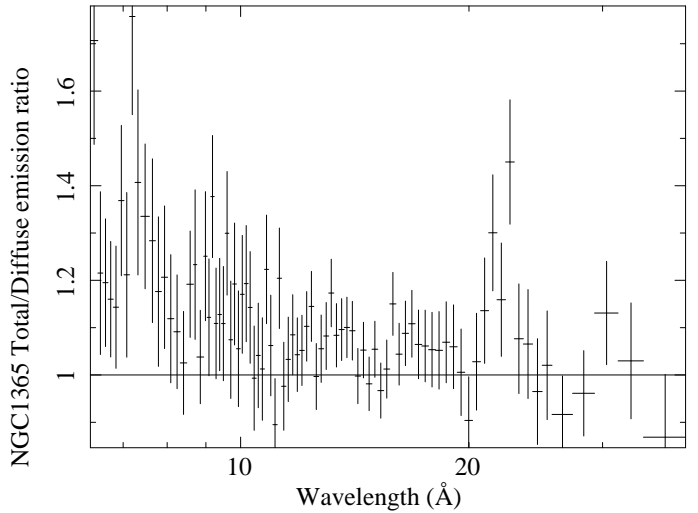}
\caption{Spectral ratio between the total NGC~1365 ACIS
spectrum and the the ACIS spectrum obtained
excising 1$\arcsec$ around the nucleus.
}
\label{fig7}
\end{figure}

An important additional information instrumental to the interpretation
of the RGS spectra comes from ACIS spatially-resolved spectroscopy.
In Fig.~\ref{fig7} we compare the best-fit
total ACIS spectrum with a
spectrum extracted from the same area once a 1\arcsec\, region
around the nucleus is excised from the spectral accumulation.
The main deviations occur once again at the wavelength of the O{\sc vii}
He-$\alpha$ triplet and at short wavelengths. The former
evidence suggests the the bulk of the photoionised plasma traced
by the He-$\alpha$ emission lines is located within the
nucleus unresolved by {\it Chandra}. The latter piece of
evidence suggests that the overall temperature of the
collisionally ionised diffuse emission increases approaching
to the centre. However, formal fits with a photoelectrically
absorbed {\tt apec} component to the two spectra confirms only
marginally this hint
(Tab.~\ref{tab4}). Only $\approxlt$10\% of the total
\begin{table}
\caption{Best-fit spectral parameters when a
simple photoelectrically absorbed collisionally ionised
plasma model
is applied to the NGC~1365 ACIS ``total'' ({\it left column}),
and ``total with
AGN excised'' (``No AGN'';
{\it right column}) spectra }
\label{tab4}
\begin{center}
\begin{tabular}{lcc} \hline \hline
& ``Total'' & ``No AGN'' \\ \hline
$N_H$ (10$^{20}$~cm$^{-2}$) & $9.4 \pm^{0.9}_{0.8}$ & $8.8 \pm 1.0$ \\
$kT$ (eV) & $620 \pm 20$ & $609 \pm^{19}_{20}$ \\
$Flux$$^a$ & $6.2 \pm 1.3$ & $5.7 \pm 1.1$ \\
$\chi^2$/~dof & 77.5/86 & 72.6/73 \\ 
\hline \hline
\end{tabular}
\end{center}

\noindent
$^a$in the 0.3--2~keV band in units of $10^{-13}$~erg~cm$^{-2}$~s$^{-1}$
\end{table}
soft X-ray emission is associated in this galaxy to the
unresolved nucleus, by contrast to what typically observed
on AGN-dominated Seyfert~2 galaxies (\cite{bianchi06}).

\section{Discussion}

NGC~1365 is the brightest Seyfert~2 known, whose soft X-ray
spectrum is energetically dominated by emission from a collisionally
ionised plasma. This component is most likely associated to
intense episodes of star formation occurring in its nuclear environment.
The analysis of the RGS spectrum presented in this paper has
provided the first proof of the statement above, while at
the same time discovering the signatures of photoionised plasma.
NGC~1365 is therefore also in X-rays a privileged laboratory to study
the connection between nuclear activity and nuclear star formation.

In this section, we put the above results in the overall context
of the soft X-ray emission of Seyfert~2 galaxies.

\subsection{NGC~1365: the paradigm of soft X-ray weak obscured AGN}

In none of the four good quality
soft X-ray high-resolution spectra of nearby AGN\footnote{NGC~1068,
Mkn~3, the Circinus Galaxy and NGC~4151} the
contribution of collisionally ionised plasma is significant. Whenever
good quality spectra can be measured, they appear dominated by
photoionisation, most likely due to the AGN, whose primary emission
illuminates the NLR gas through absorption-free lines-of-sight
(\cite{kinkhabwala02, sako00, sambruna01, armentrout07}). In NGC~1068,
by far the brightest obscured AGN of the soft X-ray sky, the
contribution of a collisionally ionised plasma is constrained to be
$\approxlt 10\%$ of the integrated 0.3--2~keV flux (\cite{brinkman02}).

\begin{figure}
\includegraphics[height=9cm, angle=90]{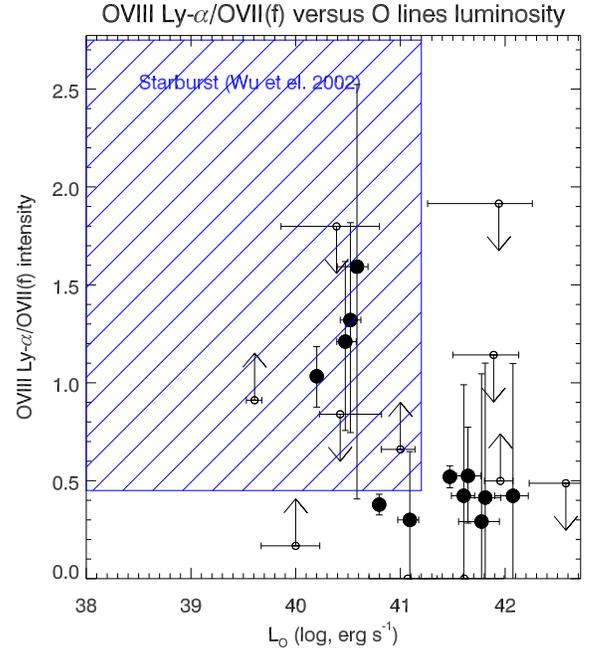}
\caption{Ratio between the O{\sc viii} Ly-$\alpha$ and
the forbidden component of the O{\sc vii} He-$\alpha$
as a function of the integrated luminosity in oxygen
lines for the Seyfert~2 galaxies of the CIELO
sample (GB07). The {\it dashed blue} area
indicate the locus occupied by a control sample
of starburst galaxies extracted from Wu et al. (2002).
{\it Filled} ({\it Empty}) circles indicate
measurements (upper/lower limits).
NGC~1365 is the {\it filled circle}
with the lowest $L_O$ value.
}
\label{fig8}
\end{figure}

Are the soft X-ray spectra of bright Seyfert~2 galaxies representative of the
whole parent population of nearby obscured AGN? GB07
showed that the distribution of line ratios in their sample of
69 Seyfert~2s observed by the XMM-Newton RGS is fairly uniform
with only a few exceptions, NGC~1365 being the most significant of
them. However,
the statistical quality of line measurements in each individual RGS spectrum
is rather low, except for a few lines in a few sources. Nonetheless,
GB07 discovered an empirical criterion, which allows us to separate {\it on
a statistical basis} the soft X-ray spectra of the CIELO Seyfert~2
galaxies from those of a control sample of starburst galaxies: the
latter occupy a specific region in a plane where the integrated
luminosity in oxygen lines and the ratio between the intensity of
the O{\sc viii} Ly-${\alpha}$ and the forbidden component of the
O{\sc vii} He-${\alpha}$ are plotted (Fig.~6 in their paper). We
present a slightly clearer version of that Figure in Fig.~\ref{fig8},
based on a re-analysis of 92 Seyfert~2 galaxies with the same method and
procedure described in GB07. Most of CIELO objects are characterised
by high oxygen luminosities ($L_O \approxgt 5 \times 10^{40}$~erg~s$^{-1}$)
and low O{\sc viii} versus O{\sc vii} ratio ($\le 0.5$). The
starburst control sample of Wu et al. (2002) nicely occupies a
complementary region in this plot.

There are a few Seyfert~2 outliers, {\it i.e.} objects characterised by
low oxygen luminosities and high O{\sc viii} versus O{\sc vii} ratios.
They are: NGC~1365, NGC~4303, NGC~5506, and NGC~7582. Apart from
NGC~1365, NGC~4303 and NGC~7582 are well known to host
intense episode of star formation close to the nucleus
(\cite{bianchi07} and references therein), which in NGC~4303 dominate
the UV and soft X-ray energy output (\cite{jimenezbailon03}).
It is therefore tempting to identify these outliers with sources, whose 
soft X-ray spectrum is substantially ``contaminated'' by collisionally
ionised optically thin emission associated to star formation.

One may try and make another step forward by adding together the information
from all those spectra, whose individual data quality is not enough to
warrant significant line detection.
We do that by stacking together rest frame spectra of
CIELO sources with total background counts in the range between
50 and 500.
In Tab.~\ref{tab6}
\begin{table}
\caption{Emission line intensities
in stacked RGS spectra of oxygen
luminosity selected, low counts CIELO sources.}
\label{tab6}
\begin{center}
\begin{tabular}{lcc} \hline \hline
Transition & Low oxygen-L$^a$ & High oxygen-L$^a$ \\ \hline
Fe {\sc xviii} 3$^s$2$^p$ & $0.47 \pm 0.26$ & $<0.14$ \\
Fe {\sc xvii} 3$^d$2$^s$ & $0.67 \pm 0.34$ & $<0.20$ \\
Fe {\sc xvii} 3$^s$2$^p$ & $1.00 \pm 0.50$ & $<0.23$ \\
Fe {\sc xxi} 3$^d$2$^p$ & $0.76 \pm 0.43$ & $<0.95$ \\
Ne {\sc ix} He-$\alpha$ (r) & $0.57 \pm 0.32$ & $0.17 \pm 0.10$ \\
Ne {\sc ix} He-$\alpha$ (i) & $<0.20$ & $<0.09$ \\
Ne {\sc ix} He-$\alpha$ (f) & $0.24 \pm 0.22$ & $0.21 \pm 0.10$ \\
N {\sc vi} He-$\alpha$ (r) & $<0.85$ & $<0.30$ \\
N {\sc vi} He-$\alpha$ (i) & $0.70 \pm 0.47$ & $0.20 \pm 0.17$ \\
N {\sc vi} He-$\alpha$ (f) & $0.93 \pm 0.56$ & $0.35 \pm 0.18$ \\
N {\sc vii} Ly-$\alpha$ & $0.72 \pm 0.44$ & $<0.17$ \\
O {\sc viii} Ly-$\alpha$ & $1.25 \pm 0.58$ & $0.48 \pm 0.14$ \\
O {\sc viii} Ly-$\beta$ & $0.44 \pm 0.26$ & $0.48 \pm 0.14$ \\
O {\sc vii} He-$\alpha$ (r) & $0.84 \pm 0.50$ & $0.52 \pm 0.20$ \\
O {\sc vii} He-$\alpha$ (i) & $<0.54$ & $<0.24$ \\
O {\sc vii} He-$\alpha$ (f) & $1.00 \pm 0.41$ & $1.00 \pm 0.20$ \\
\hline \hline
\end{tabular}
\end{center}

\noindent
$^a$normalised to the O{\sc vii} He-$\alpha$ (f)

\end{table}
we show
the normalised intensity of the emission lines
detected in the stacked spectra for objects with total
integrated oxygen lines luminosity smaller and higher
then $L_O \sim 10^{41}$~erg~s$^{-1}$.
Errors in this Table are at 1-$\sigma$.
Low-oxygen luminosity sources exhibit: a) stronger Fe-L lines; 
b) stronger recombination and intercombination transitions
in He-$\alpha$ triplets of Ne{\sc xi} and N{\sc vi};
c) stronger Ly-$\alpha$ transitions of O{\sc viii} and
Ne{\sc vii}
All the above trends are consistent with a stronger role played by
collisionally ionised plasmas in low-oxygen luminosity sources.

\subsection{BLR shielding of the ionising continuum?}

CIELO Seyfert~2 which are
under-luminous in oxygen lines are not necessarily {\it intrinsically}
low-luminous. If we consider those galaxies,
which have a measurement of the O{\sc viii} Ly-$\alpha$ versus
O{\sc vii}(f) intensity ratio in Fig.~\ref{fig8}, and exclude
NGC~4303, where the presence of an AGN is still under debate
(\cite{jimenezbailon03}):
the absorption-corrected 2--10~keV luminosity is $2.5 \times
10^{42}$~erg~s$^{-1}$ in NGC~1365 (\cite{risaliti09a});
1.5--3.2$\times 10^{42}$~erg~s$^{-1}$ in NGC~7582
(\cite{bianchi09a}), and 4.5--9.1$\times 10^{42}$~erg~s$^{-1}$ in
NGC~5506 (\cite{bianchi03}). NGC~1365 and NGC~7582 share another remarkable
observational property: both exhibit extreme
(by one order of magnitude) variability of
the X-ray absorbing column density on timescales as short
as 10~hours (\cite{risaliti09b}) and $\approxlt$~1~day
(\cite{bianchi09a}), respectively. Such rapid transitions
from Compton-thin (transmission-dominated) to Compton-thick
(reflection-dominated) states can be explained only if
the X-ray absorbers are located at distances from the
black hole comparable to that of the broad line region (BLR). It is
intriguing to speculate that the relative weakness of
the AGN photoionisation contribution in these objects, as
well as the overall weakness of their soft X-ray emission lines,
are due to shielding of the ionising high-energy AGN
emission by optically thick matter in the BLR.
In this respect, however, NGC~5506 does not seem to match
this scenario. Large (by a factor of about 2) historical
variation of the X-ray flux are accompanied by a very modest
variability of the absorbing column density
(Guainazzi et al.,
in preparation). The column
density of NGC~5506 (\cite{bianchi03})
is also significantly smaller than measured even
in the Compton-thinnest states of NGC~1365 and NGC~7582
($N_H \simeq$2.5$\times 10^{22}$~cm$^{-2}$). It is still
possible that such a small column density is associated
to matter in the host galaxy rather than in the nuclear
environment (\cite{piconcelli07,matt03}), and that the
ENLR gas is looking at the AGN trough a different
optical path where obscuration of the active nucleus is
substantially high. As already pointed out by
several authors, we are only now starting to glimpse
the true complexity of the gas and dust distribution
in the innermost parsec around supermassive black holes
(\cite{elvis00,risaliti05b,elitzur06}).

\subsection{Properties of the photoionised plasma in NGC~1365}

It is impossible to significantly constraint the properties of
the photoionised plasma responsible for the bulk of the
forbidden component of He-like triplets in the RGS spectrum
of NGC~1365. With the assumed SED: $\log(U) = 1.6 \pm^{0.3}_{0.4}$
(1$\sigma$),
$n_e \approxlt 10^{10}$cm$^{-3}$, and
$N_H \approxgt 10^{22}$cm$^{-2}$. From the
best-fit {\it CLOUDY} solution (Sect.~4.2), and using the
definition of ionisation parameter, we estimate that the inner
side of the photoionised nebula is constrained to be located
$\approxgt 0.75$~pc from the source of the ionising photons.
This is still consistent with a location of this gas within the
BLR. However, it is not consistent with
the soft X-ray photoionised gas being the same responsible for
the highly ionised, outflowing iron absorption lines discovered
by {\it XMM-Newton} (\cite{risaliti05a}), and probably
associated to a disk outflow. The width of the 
oxygen RRC features constrains the gas temperature to be
$\sim$10$^5$~K, of the same order of
magnitude as the temperature measured in NGC~1068 (\cite{kinkhabwala02})
or other CIELO AGN (GB07).
The only He-like ion, for whose He-$\alpha$ triplet meaningful
constraints on the diagnostic parameters $G$ and $R$ (\cite{porquet00})
can be
derived is O{\sc vii}.
The latter is primarily sensitive to density.
$R \equiv f/i = 3.6 \pm 1.2$ implies $\log (n_e) \approxlt 10.5$,
in agreement with the results of the comparison between
{\sc Cloudy} simulations and the RGS line spectrum.
On the other hand, $G \equiv (f+i)/r > 4.2$ translates into a
loose constraints on the photoionised plasma temperature,
$\approxlt 9 \times 10^5$~K, again in agreement
with the RRC-based diagnostic.

\subsection{Properties of the collisionally ionised plasma in NGC~1365}

A formally good fit to the RGS spectrum of NGC~1365
requires two optically thin, collisionally ionised thermal
components with temperatures $\simeq$300~eV and  $\simeq$640~eV.
These values should not be interpreted too literally, though.
The best-fit temperature is mainly driven by the fit of
the emission line strength and position. The tightest
constraints are provided by the strongest lines, which
are mostly emitted at $\simeq$0.3~keV ({\it e.g}: oxygen and
neon lines) and $\simeq$0.7~keV (Mg lines). Gas at
various other temperatures may be also present, however it
could not get the same weight in the fitting process because it
does not emit as strongly in those lines. It is likely that
this results should be interpreted as a distribution of
temperature over the nominal best-fit range.

Our analysis qualitatively confirms the results by
Wang et al. (2009)
that the gas in the circumnuclear environment of
NGC~1365 has an overall very low metallicity.
However, RGS lacks spatial resolution
matching the complex morphology of metal abundances
in the NGC~1365 nuclear ambient gas. Besides the
overall metal under-abundance, regions of silicon and
iron {\it over}-abundance were discovered by {\it Chandra}
(\cite{wang09}). Moreover, the high Equivalent Width
of the relativistically broadened iron K$_{\alpha}$ line
measured by the XMM-Newton EPIC (\cite{risaliti09a})
is also consistent with a factor of about three
iron overabundance.

Gas in the spiral arm of the galaxy should be have an
even lower metallicity. At the same time, NGC~1365 has a
high number of ULX in its disk (\cite{soria09}). These
pieces of evidence
are consistent with a scenario, whereby ULXs are
black hole remnants of very metal-poor ($Z \le 0.1 Z_{\odot}$)
stars, which little mass in their winds
($\dot{m} \sim Z^{0.7}$).

\begin{figure}
\includegraphics[height=9cm]{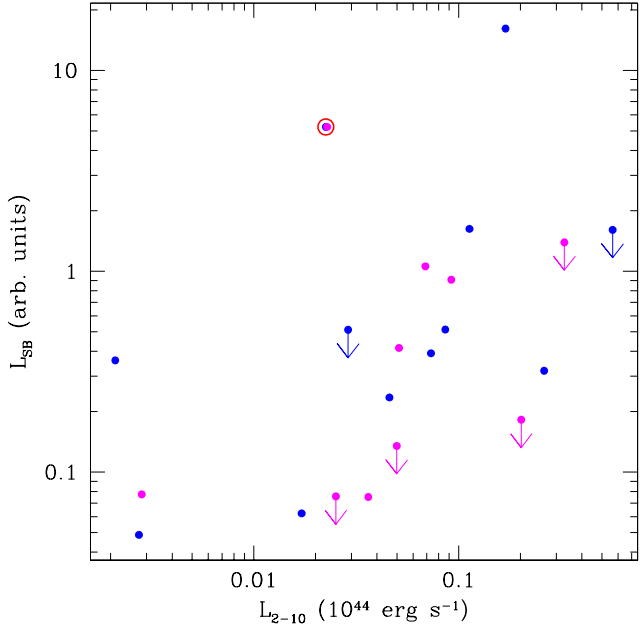}
\caption{6.2$\mu$m PAH feature flux (in arbitrary units)
versus the 2--10~keV absorption corrected luminosity
in a sample of obscured ({\it grey}) and
unobscured ({\it black}) AGN. NGC~1365 is the
dot surrounded by an {\it empty circle}.
}
\label{fig12}
\end{figure}

Does NGC~1365 host a more luminous starburst when
compared with obscured AGN of comparable intrinsic AGN
power? In Fig.~\ref{fig12} we plot the flux of the
PAH feature at 6.2$\mu$ as measured by {\it Spitzer}
as a function of the unobscured 2--10~keV luminosity
as measured by CAIXA (\cite{bianchi09b})
in a sample of obscured and
unobscured local Universe AGN.
PAH features are a strong indicator of star formation
in active galaxies (see, {\it e.g.}, \cite{lutz08} and
references therein). NGC~1365 is the second brightest
PAH AGN in the sample after NGC~7469; the latter is well
known to exhibit an extremely bright starburst at distances
between tens to about 10$^3$ of parsecs from the AGN
(\cite{diazsantos07}).

\section{Conclusions and summary} 

In this paper we primarily discuss the 0.5~Ms RGS spectrum of
the Seyfert~2 galaxy NGC~1365. The main results of our analysis
can be summarised as follows:

\begin{itemize}

\item no variability either in the overall spectral shape or in
the line intensity is detected when comparing observations distant
up to five years apart

\item the spectrum is dominated by emission lines from
optically thin, collisionally ionised plasma, plus a strong
continuum primarily in the 10--20 \AA\, range.
Formally, a good fit is obtained with two components,
with temperatures $\simeq$300~eV and  $\simeq$640~eV,
the latter seen through a photoelectric absorption column
density $\simeq 3 \times 10^{21}$~cm$^{-2}$.
These results are in broad agreement with the range of
temperatures and column densities measured by
spatially resolved moderate resolution
spectroscopy of this galaxy with
{\it Chandra} (\cite{wang09})

\item the RGS spectrum confirms the suggestion by {\it Chandra}
that on the average sub-solar metallicities are required to
fit the data, although the unknown level of contamination by
individual point-like sources makes the exact measurements of
metallicities from the RGS spectrum uncertain

\item a residual contribution by a photoionised plasma is
required to fit the forbidden component of He-like transitions
of oxygen and nitrogen, as well as the intercombination
component of O{\sc vii}.
The physical parameters of the photoionised plasma can be
only loosely constrained:  $\log(U) \simeq$1.6,
$n_e \approxlt 10^{10}$cm$^{-3}$, and
$N_H \approxgt 10^{22}$cm$^{-2}$, and correspond to a plasma at a distance
$\approxgt 0.75$~pc. The detection of oxygen RRC features
suggests a photoionised temperature $\sim$10$^5$~K,
in agreement with values typically observed in AGN-dominated Seyfert~2
galaxies.

\end{itemize}

Despite its comparably low luminosity in soft X-ray lines
with respect to AGN-photoionisation dominated Seyfert~2
galaxies, NGC~1365 is not {\it intrinsically} X-ray weak.
We speculate that the soft X-ray spectrum
NGC~1365 could appear as starburst-dominated
due to shielding of the primary AGN ionising continuum by
optically thick matter associated to the BLR, a scenario
which may apply to other AGN (such as NGC~7582), which exhibit
extreme and rapid variability of the absorber column density.
Good-quality soft X-ray spectra of ``changing-look'' AGN
would be crucial to test this hypothesis. 
Alternatively,
NGC~1365 could host a remarkably luminous nuclear starburst when
compared to its AGN accretion power, as suggested by the
luminous PAH features detected in its {\it Spitzer}
spectra.

\begin{acknowledgements}

Based on observations obtained with XMM-Newton, an ESA science mission
with instruments and contributions directly funded by ESA Member
States and NASA
This research has made use of
data obtained through the High Energy Astrophysics Science Archive
Research Centre Online Service, provided by the NASA/Goddard Space
Flight Centre and of the NASA/IPAC Extragalactic Database (NED) which
is operated by the Jet Propulsion Laboratory, California Institute of
Technology, under contract with the National Aeronautics and Space
Administration. A careful reading of the original
manuscript by an anonymous referee is gratefully acknowledged.

\end{acknowledgements}


\begin{thebibliography}{}

\bibitem[Anders \& Grevesse 1989]{anders89} Anders E. \& Grevesse N., 1989, Geochimica et Cosmochimica Acta 53, 197

\bibitem[Armentrout et al. 2007]{armentrout07} Armentrout B.K., Kramer S.B., Turner T.J., 2007, ApJ, 665, 237

\bibitem[Arnaud 1996]{arnaud96} Arnaud, K.A., 1996, Astronomical Data Analysis Software and Systems V, eds. Jacoby G. and Barnes J., 17, ASP Conf. Series volume 101.

\bibitem[Axon et al. 1998]{axon98} Axon D.J., Marconi A., Capetti A., Macchetto F.D., Schreier E., Robinson A., 1998, ApJ, 496, L75

\bibitem[Bennert et al. 2006]{bennert06} Bennert N., Jungwiert B., Komossa S., Haas M., Chini R., 2006, A\&A, 456, 953

\bibitem[Bianchi et al. 2003]{bianchi03} Bianchi S., Balestra I., Matt G., Guainazzi M., Perola G.C., 2003, A\&A, 402, 141

\bibitem[Bianchi et al. 2007]{bianchi07} Bianchi S., Chiaberge M., Piconcelli E., Guainazzi M., 2007, MNRAS, 374, 697

\bibitem[Bianchi et al. 2006]{bianchi06} Bianchi S., Guainazzi M., Chiaberge M., 2006, A\&A, 448, 499

\bibitem[Bianchi et al. 2009a]{bianchi09a} Bianchi S., Piconcelli E., Chiaberge M., Jim\'enez-Bail\'on E., Matt G., Fiore F., 2009a, ApJ, 695, 781

\bibitem[Bianchi et al. 2009b]{bianchi09b} Bianchi S., Guainazzi M., Matt G., Fonseca Bonilla N., Ponti G., A\&A, 2009b, 495, 421

\bibitem[Borkowski et al. 2001]{borkowski01} Borkowski K.J., Lyerly W.J., Reynolds S.P., 2001, ApJ, 548, 820

\bibitem[Bradley et al. 2004]{bradley04} Bradley L.D., Kaiser M.E., Baan W.A., 2004, ApJ, 603, 463

\bibitem[Brinkman et al. 2002]{brinkman02} Brinkman A.C., Kaastra J.C., van der Meer R.J.L., Kinkhabwala A., Behar E., Kahn S., Pearels F.B.S., Sako M., 2002, A\&A, 396, 761

\bibitem[Capetti et al. 1996]{capetti96} Capetti A., Axon D.J., Macchetto F., Sparks W.B., Boksenberg A., 1996, ApJ, 469, 554

\bibitem[Das et al. 2006]{das06}Das V., Crenshaw D.M., Kraemer S.B., Deo R.P., 2006, AJ, 132, 620

\bibitem[den Herder et al. 2001]{denherder01} den Herder J.,
et al., 2001, A\&A, 365, L7

\bibitem[D\'iaz-Santos et al. 2007]{diazsantos07} D\'iaz-Santos T., Alonso-Herrero A., Colina L., Ryder S.D., Knapen J.H., 2007, ApJ, 661, 149

\bibitem[Elitzur \& Shloshman 2006]{elitzur06} Elitzur M. \& Sholshman I., 2006, ApJ, 648, L101

\bibitem[Elvis 2000]{elvis00} Elvis M., 2000, ApJ, 545, 63

\bibitem[Ferland et al. 1998]{ferland98} Ferland G. J., Korista K.T., Verner D.A., Ferguson J.W., Kingdon J.B., Verner E.M., 1998, PASP, 110, 761

\bibitem[Ferland \& Osterbrock 1986]{ferland86} Ferland G.J., Osterbrock D.E., 1986, ApJ, 300, 658

\bibitem[Gabriel et al. 2003]{gabriel03} Gabriel C., Denby M., Fyfe D. J., Hoar J., Ibarra A., 2003, in ASP Conf. Ser., Vol. 314 Astronomical Data Analysis Software and Systems XIII, eds. F. Ochsenbein, M. Allen, \& D. Egret (San Francisco: ASP), 759 

\bibitem[Galliano et al. 2005]{galliano05} Galliano E., Alloin D., Pantin E., Lagage P.O., Marco O., 2005, A\&A, 438, 803

\bibitem[Galliano et al. 2008]{galliano08} Galliano E., Alloin D., Pantin E., et al., 2008, A\&A, 492, 3

\bibitem[Grimm et al. 2003]{grimm03} Grimm H.-J., Gilfanov M., Sunyaev R., 2003, MNRAS, 339, 793

\bibitem[Guainazzi \& Bianchi 2007]{guainazzi07} Guainazzi M., Bianchi S., 2007, MNRAS, 374, 1290 (GB07)

\bibitem[Hjelm \& Lindblad 1996]{hjelm96} Hjelm M., Lindblad P.O., 1996, A\&A, 305, 727

\bibitem[Kalberla et al. 2005]{kalberla05} Kalberla P.M.W., Burton W.M., Hartmann D., et al., 2005, A\&A, 440. 775

\bibitem[Kinkhabwala et al. 2002]{kinkhabwala02} Kinkhabwala A., Sako M., Behar E., et al., 2002, ApJ, 575, 732 

\bibitem[Kraemer et al. 2000]{kraemer00} Kraemer S.B., Crenshaw D.M., Hatchings J.B., et al., 2000, ApJ, 531, 278

\bibitem[Kraemer et al. 2008]{kraemer08} Kraemer S.BN., Schnitt H.R., Crenshaw D.M., 2008, ApJ, 679, 1128

\bibitem[Kristen et al. 1997]{kristen97} Kristen H., Jorsater S., Lindblad P.O., Boksenberg A., 1997, A\&A, 328, 483

\bibitem[Jim\'enez-Bail\'on et al. 2003]{jimenezbailon03} Jim\'enez-Bail\'on E., Santos--LLeo M., Mas-Hesse M., et al., 2003, ApJ, 593, 127

\bibitem[Lampton et al. 1976]{lampton76} Lampton M., Margon B., Bowyer S., 1976, ApJ, 207, 894

\bibitem[Liedahl et al. 1995]{liedahl95} Liedahl D.A., Osterheld A.L., Goldstein W.H., 1985, ApJL, 438, 115

\bibitem[Longinotti et al. 2008]{longinotti08} Longinotti A.L., Nucita A., Santos--LLeo M., Guainazzi M., 2008, A\&A, 484, L311

\bibitem[Lutz et al. 2008]{lutz08} Lutz D., Sturm E., Tacconi L.J., et al., 2008, ApJ, 684, 753

\bibitem[Maiolino \& Rieke 1995]{maiolino95} Maiolino R., Rieke G.H., 1995, ApJ, 454, 95

\bibitem[Mass-Hesse et al. 2008]{mashesse08} Mas-Hesse J.M., Ot\'i-Floranes H., Cervi\~no M., 2008, A\&A, 483, 71

\bibitem[Matt et al. 2003]{matt03} Matt G., Bianchi S., Guainazzi M., et al., 2003, A\&A, 399, 519

\bibitem[Mewe et al. 1985]{mewe85} Mewe R., Gronenschild E.H.B.M., van der Oord G.H.J., 1985, A\&AS, 62, 197

\bibitem[Osterbrock 1989]{osterbrock89} Osterbrock D.E., 1989, {\it Astrophysics of Gasous Nebulae and Active Galactic Nuclei} (University Science Books: Mill Valley)

\bibitem[Persic \& Raphaeli 2002]{persic02} Persic M., Raphaeli Y., 2002, A\&A, 382, 843

\bibitem[Piconcelli et al. 2007]{piconcelli07} Piconcelli E., Bianchi S., Guainazzi M., Fiore F., Chiaberge M., 2007, A\&A, 466, 855

\bibitem[Plucinsky et al. 2008]{plucinsky08} Plucinsky P., Haberl F., Dewey D., et al., 2008, SPIE, 7011, 68

\bibitem[Pogge 1988]{pogge88} Pogge R.W., 1988, ApJ, 328, 519

\bibitem[Porquet \& Dubau 2000]{porquet00} Porquet D., Dubau J., 2000, A\&AS, 143, 495

\bibitem[Risaliti et al. 2005a]{risaliti05a} Risaliti G., Bianchi S., Matt G., et al., 2005a, ApJ, 630, L129

\bibitem[Risaliti et al. 2005b]{risaliti05b} Risaliti G., Elvis M., Fabbiano G., Baldi A., Zezas A., 2005b, ApJ, 623, L93

\bibitem[Risaliti et al. 2007]{risaliti07} Risaliti G., Elvis M., Fabbiano G., Baldi A., Zezas A., Salvati M., 2007, ApJ, 659, L111

\bibitem[Risaliti et al. 2000]{risaliti00} Risaliti G., Maiolino R., Bassani L., 2000, A\&A 356, 33

\bibitem[Risaliti et al. 2009a]{risaliti09a} Risaliti G., Miniutti G., Fabbiano G., et al., 2009a, ApJ, 696, 160

\bibitem[Risaliti et al. 2009b]{risaliti09b} Risaliti G., Salvati M., Elvis M., et al., 2009b, MNRAS, 393, L1

\bibitem[Sakamoto et al. 2007]{sakamoto07} Sakamoto K., Ho P.T., Mao R.-Q., Matsushita S., Peck A.B., 2007, ApJ, 654, 782

\bibitem[Sako et al. 2000]{sako00} Sako M., Kahn S.M., Paerels F., Liedahl D.A., 2000, ApJL 543, L115

\bibitem[Sambruna et al. 2001]{sambruna01} Sambruna R., Netzer H., Kaspi S., Brandt W.N., Chartas G., Garmire G.P., Nousek J.A., Weaver K.A., 2001, ApJ, 546, L13

\bibitem[Sandqvist et al. 1995]{sandqvist95} Sandqvist A., Joersaeter S., Lindblad P.O., 1995, A\&A, 295, 585

\bibitem[Silbermann et al. 1999]{silbermann99} Silbermann N.A., Harding P., Ferrarese L., et al., ApJ, 515, 1

\bibitem[Soria et al. 2009]{soria09} Soria R., Risaliti G., Elvis M., Fabbiano G., Bianchi S., Kuncic Z., 2009, ApJ, 695, 1614

\bibitem[Smith et al. 2001]{smith01} Smith R.K., Brickhouse N.S., Liedahl D.A., Raymond J.C., 2001, ApJ, 556, 91

\bibitem[Strateva \& Komossa 2008]{strateva08} Strateva I.V., Komossa S., 2009, ApJ, 692, 443

\bibitem[Tadhunter \& Tsvetanov 1989]{tadhunter89} Tadhunter C., Tsvetanov Z., 1989, Nature, 341, 422

\bibitem[Turner et al. 1993]{turner93} Turner T.J., Urry C.M., Mushotzky R.F., 1993, ApJ, 418, 653

\bibitem[Ulvestad \& Ho 2001]{ulvestad01} Ulvestad J.S., Ho L.C., 2001, ApJ, 558, 561

\bibitem[Veron et al. 1980]{veron80} Veron P., Lindblad P.O., Zuiderwijk E.J., Veron M.P., Adam G., 1980, A\&A, 87, 245

\bibitem[Viegas-Aldrovandi \& Contini 1989]{viegasaldrovandi89} Viegas-Aldrovandi S.M., Contini M., 1989, ApJ, 339, 689

\bibitem[Veilleux et al. 2003]{veilleux03} Veilleux S., Shopbell P.L., Rupke D.S., Bland-Hawthorn J., Cecil G., 2003, AJ, 126, 2185

\bibitem[Young et al. 2001]{young01} Young A.J., Wilson A.S., Shopbell P.L., 2001, ApJ, 556, 6

\bibitem[Wang et al. 2009]{wang09} Wang J., Fabbiano G., Elvis M., et al. 2009, ApJ, 694, 718

\bibitem[Whittle 1985]{whittle85} Whittle M., 1985, MNRAS, 213, 33

\bibitem[Whittle 1992]{whittle92} Whittle M., 1992, ApJS, 79, 49

\bibitem[Wilson et al. 2000]{wilson00} Wilson A.S., Shopbell P.L., Simpson C., Storchi-Bergmann T., Barbosa F.K.B., Warm M.J., 2000, AJ, 120, 1325

\bibitem[Wilson \& Willis 1980]{wilson80} Wilson A.S., Willis A.G., 1980, ApJ, 240, 429

\bibitem[Wu et al. 2002]{wu02} Wu W., Clayton G.C., Gordon K.D., Misselt K.A., Smoth T.L., Calzetti D., 2002, ApJS, 143, 377

\end{thebibliography}
\end{document}